\documentclass[11pt,a4paper]{article}

\usepackage[utf8]{inputenc}
\usepackage[english]{babel}
\usepackage{amsmath}
\usepackage{amsfonts}
\usepackage{amssymb}
\usepackage{graphicx}
\usepackage[labelfont=bf]{caption}
\usepackage{tikz}
\usetikzlibrary{shapes,arrows,shadows,decorations.pathreplacing,angles,quotes}
\usepackage{subcaption}
\usepackage{authblk}
\usepackage[toc]{appendix}
\usepackage{xcolor}
\usepackage[normalem]{ulem}
\usepackage{authblk}
\usepackage[round]{natbib}
\usepackage{verbatim}

\usepackage[margin=2cm]{geometry}
\usepackage[sc]{mathpazo}
\usepackage{url}

\graphicspath{{Images/}}
\newcommand{\dtone}{\ensuremath{\Delta t_1}}
\newcommand{\dttwo}{\ensuremath{\Delta t_2}}

\title{The market nanostructure origin\\ of asset price time reversal asymmetry}

\author[1]{Marcus Cordi}
\author[1]{Damien Challet\footnote{Corresponding author: damien.challet\symbol{64}centralesupelec.fr}}
\author[2]{Serge Kassibrakis}

\affil[1]{Université Paris-Saclay, CentraleSupélec, Mathématiques et Informatique pour la Complexité et les Systèmes, 91190, Gif-sur-Yvette, France.}
\affil[2]{Swissquote Bank SA, chemin de la Crétaux 33, 1196 Gland, Switzerland}

\begin{document}
\maketitle
\begin{abstract}
We introduce a method to infer lead-lag networks between the states of elements of complex systems, determined at different timescales. As such networks encode a causal structure of a system, inferring lead-lag networks for many pairs of timescales provides a global picture of the mutual influence between timescales.
We apply our method to two trader-resolved FX data sets and document strong and complex asymmetric influence of timescales on the structure of lead-lag networks. This asymmetry extends to the propagation of trader activity between timescales. For both retail and institutional traders, we find that historical activity over longer timescales has a greater correlation with future activity over shorter timescales (Zumbach effect), for sufficiently large timescales both in the past and future (about one hour for retail traders and two hours for institutional traders); remarkably the effect is opposite for  smaller timescales, and much weaker for retail traders.
\begin{description}
\item [{Keywords:}]  Trader activity, lead-lag networks, timescales, volatility structure.{\small \par}
\end{description}
\end{abstract}

\newpage

\section{Introduction}
The collective behavior of investors  plays a major part in shaping the complexity of price dynamics. A major challenge in the analysis and modelling of market dynamics comes from the significant heterogeneity of market participants, particularly with respect to their activity rate and feedback speed. 
Most agent-based models of financial markets omit timescale heterogeneity, usually focusing on strategy heterogeneity (fundamentalists, trend-followers or noise traders) and the way they learn to use them (see \citet{hommes2006heterogeneous} for a review). Notable exceptions include \citet{marsili2002colored,mosetti2006minority,kroujiline2016forecasting}.

The typical time-horizon of trader activity ranges from a fraction of a second to a few months \citep{dacorogna1998modelling,zumbach2009time}. 
A fundamental challenge is thus to characterize the causal structure of market activity across timescales.
In other words, is there a hierarchical (or more complex) structure on which trading activity propagates?
Since trader-resolved data is hard to obtain, past works focused on price dynamics and volatility propagation between timescales. Intuitively, the price dynamics should reflect in some way heterogeneous trader time horizons (see e.g. \citet{muller1993fractals}). 
Early works exploit the intuitive analogy between turbulent flows and price changes \citep{ghashghaie1996turbulent}, and simple cascade models of the price dynamics have been proposed \citep{lux2001turbulence}. 
Heterogeneous trader timescales may also explain why multiscale GARCH models are generally much better than plain GARCH ones (see e.g. \citet{lynch2003market,borland2005multi,chicheportiche2014fine}). 
In particular, \citet{muller1997volatilities} argue that since coarsely-defined volatility predicts finely-defined volatility significantly better than the other way around, the behavior of long-term traders should influence the behavior of short-term traders.

The above discussion implicitly assumes that prices are Time Reversal Asymmetric (TRA). 
\citet{zumbach2001heterogeneous}, and \citet{zumbach2009time} show that financial time-series are indeed significantly asymmetric with respect to the reversal of the arrow of time by using two different kinds of time resolution in the future and in the past. 
While classical models of price and volatility dynamics are not TRA, GARCH processes that incorporate price returns defined over several timescales are TRA \citep{zumbach2001heterogeneous,zumbach2009time,chicheportiche2014fine}. 
Ultimately, volatility TRA comes at least in part from trader activity TRA, which motivates the present work.
  
\section{Methods}

The method we introduce here is generic and may be applied to other types of data in which the state of agents is known. The crucial assumption is that the state of these agents can be summarized by a discrete variable that is restricted to a small set of possible values (typically much smaller than the number agents). In the case of traders, the simplest natural set of states over a given time interval are buy, sell, neutral, and no activity. While this choice discards any information about the volume traded (which significantly differs between traders), it is computationally faster. One could extend this approach in order to keep information about volumes while still using discrete states by grouping buy and sell trades according to a few quantiles of the volumes traded over a given timescale, determined in each calibration window, which would keep the number of possible states acceptably low.

Because trader activity is very heterogeneous, special care has to be taken. We propose here to infer lead-lag networks (or equivalently, causality) of the state of traders determined over two different timescales with robust methods and then to test for significant differences between the average reciprocal influences of these timescales. Instead of working with all the traders at once, we first cluster traders into groups for each timescale and focus on the causality between these groups. This reduces the complexity of the problem by a factor 10-100, which is welcome as our computations take a few days on 72 modern CPU cores for each data set and for each calibration window length. In addition, as explained below, grouping traders according to the similarity of their buy/sell patterns leads to a consistent aggregation of activity. 

Groups of traders are determined with Statistically Validated Networks (SVNs); SVNs were introduced by \citet{tumminello2011statistically} and have been applied e.g. to mobile communication networks \citep{li2014statistically}, clusters of orthologous genes, and the relationship between actors and movies \citep{tumminello2011statistically}. 
They were then used to cluster Finnish investors \citep{tumminello2012identification} and more recently to understand their long-term ecology \citep{musciotto2018long}.
The main idea is that a group of similar traders should act in a similar way, for instance because they use the same strategy or news source. The first step is to define networks of interaction according to the significant pairwise synchronization between the actions of traders. Groups of traders are then identified with community detection methods, a family of clustering methods designed for networks. Crucially, since the actions of all the members of a group are remarkably similar, the aggregate action of the group is representative of the action of each of its members: if all the agents belonging in a group typically buy at the same time, the state of the group will naturally be ``buy'': aggregation is consistent.

SVNs rely on time coarsening at a given timescale (e.g. one day in \citet{tumminello2012identification}  or one hour in \citet{challet2018trader}). While only these two timescales have been used in the past, which timescale to choose is not obvious, all the more since traders have widely different activity rates and since a relevant timescales for retail traders may differ from the one for institutional traders. We fill this gap by a systematic study of the dependence of SVNs on the coarsening timescale.
As we shall see below, the answer depends on the type of traders (retail or institutional) and most probably on the clientele of a broker.

\citet{challet2018trader} introduced  Lead-Lag SVNs (LL-SVNs) to infer causality networks between the states of agents in complex systems and applied them to trader-resolved data. 
In brokerage data, the persistence in LL-SVNs is large enough to make it possible to predict the sign of the order flow and of the change of volume-weighted average price  (VWAP) at which the clients of a broker trade, at least over the next hour. 
A probable reason why these lead-lag networks exist and persist is that investors consistently react with different speeds to common information \citep{boudoukh1994tale,jegadeesh1995overreaction}. However, this method is not able to detect TRA, as the timescales of the leading and lagging groups must be the same ones, whereas TRA detection with this kind of approach requires different timescales \citep{zumbach2009time}.

Here, we extend the LL-SVN method to infer causality networks between agent states determined at two different timescales. When applied to many pairs of timescales, this method provides a fine picture of how information propagates in a complex system over a large variety of timescales. We also discuss how the TRA of the activity of the two types of traders in our data set compares with that of the volatility, i.e., how to relate macroscopic price properties to the decisions of traders, which we propose to call market nanostructure.

\subsection{Statistically Validated Networks}\label{sec:SVN}
Assume that one has $N$ time series, e.g.\ the history of the transactions of $N$ traders. When the time stamp resolution of these data is precise enough, these time series may be asynchronous. This is why one transforms them into synchronous time series by coarsening time into regularly-spaced time slices:
\begin{enumerate}
\item one chooses a time resolution $\Delta t$ (e.g. 5 minutes, 1 hour, etc.);
\item each time series is split into non-overlapping slices of length $\Delta t$;
\item the state of each agent is determined in each slice (see below).
\end{enumerate}

Following \cite{tumminello2011statistically}, we define the state of each trader $i$ and each time slice $t$ as mostly buying (+1), mostly selling (-1), neutral (0) and inactive (NA) according to the imbalance ratio of its buy and sell volume during that time slice. Adopting the usual convention that a buy trade has a positive volume $(\omega>0)$ and a sell trade correspond to a negative volume $\omega<0$, one defines
\begin{equation}
\rho_i(t)=\frac{v_i(t)}{a_i(t)}
\end{equation}
where $v_i(t)$ is the total net transaction volume of trader $i$ during timeslice $t\equiv [t,t+\Delta t[$,\footnote{This notation means from $t$ to $t+\Delta t$, $t+\Delta t$ not included} i.e., $v_i(t)=\sum_{k}\omega_{i,k}(t)$ where $\omega_{i,k}(t)$ is the volume of the $k$-th transaction of trader $i$ during timeslice $t$,  and $a_i(t)$ is the turnover during this timeslice $t$, defined as $a_i(t)=\sum_{k}|\omega_{i,k}(t)|$. The state of agent $i$ during timeslice $t$ is 
\begin{equation}
  \sigma_i(t)=\begin{cases}
    +1 & \text{if $\rho_i(t)>+\rho_0$}\\
    -1 & \text{if $\rho_i(t)<-\rho_0$}\\
    0 & \text{if $\rho_i(t)<\lvert\rho_0\rvert$}\\
    \textrm{NA} & \text{if $v_i(t)=a_i(t)=0$.}
  \end{cases},
\end{equation} 
As in previous works, we use $\rho_0=0.01$, but the specific choice of this parameter does not have much influence on the results provided that it is smaller than $0.1$. 

In statistically validated networks, two agents are linked only if their level of synchronicity for a given pair of states (e.g. simultaneous ``mostly buying'' $(+1,+1)$) is statistically surprisingly large. More precisely, synchronicity is defined with respect to a null hypothesis: assuming that the states of agents are determined by Poisson processes, the probability of having a given number of simultaneous co-occurences of states is known exactly and can be used to compute a p-value.
The computation of p-values is then performed for all possible pairs of traders and all allowed state pairs  (see \cite{tumminello2012identification} for more details).

Because we wish to group agents that behave in a similar fashion, the set of relevant pairs of states is $\{(1,1),(-1,-1),(0,0)\}$. We focus here on the most active traders and hence will ignore the inactive state.
Testing all the pairs of traders for each possible state pair yields a very large number of tests, thus multiple hypothesis testing correction of the critical p-value is needed: we use the False Discovery Rate (FDR) \citep{benjamini1995controlling}, with an FDR rate set to $p_0=0.05	$. An SVN is obtained by keeping links whose p-values are smaller than the FDR-adjusted threshold (see \citet{tumminello2012identification} for more details). To summarize, the expected fraction of wrong links in our SVNs is 5\%. Note that there may be more than one link between two agents since they may be synchronous for more than one state pair (e.g., they may buy and sell synchronously).

The resulting SVN  may then be decomposed into groups (communities) by using a suitable network community detection method. We use InfoMap method \citep{rosvall2008maps}, which is one of the most efficient methods \citep{lancichinetti2009community}. Finally, as explained above, the state (mostly buying, mostly selling, etc.) of each group of traders is well defined because traders are clustered according to the similarity of their actions, thus the state of each group mirrors the actions of the traders that it includes. 

Let us introduce  more notations.  
Mathematically, one can define the state of group $g\in G$, where $G$ is the set of all groups, during timeslice $t$, by
\begin{equation}
  \sigma_g(t)=\begin{cases}
    1 & \text{if $\rho_g(t)>\rho_0$}\\
    -1 & \text{if $\rho_g(t)<-\rho_0$}\\
    0 & \text{if $\rho_g(t)<\lvert\rho_0\rvert$}\\
  \end{cases}
\end{equation} 
where $\rho_g(t)=\frac{V_g(t)}{A_g(t)}$, $V_g(t)=\sum\limits_{i\in g}v_i(t)$ is the aggregate signed volume of the traders belonging to group $g$ during timeslice $t$ and $A_g(t)=\sum\limits_{i\in g}a_i(t)$ is the aggregate absolute volume of the traders belonging to group $g$ during timeslice $t$.
Grouping traders is very efficient and significantly decreases  the dimensionality of the problem, i.e., the effective number of timeseries to track in a population of clients of a broker.

\subsection{Lead-Lag SVNs}

\citet{challet2018trader} introduce a method to infer lead-lag networks, which simply consists in applying SVNs between states of a time series and the lagged states of another time series (or possibly the same one), which yields causality networks. A link in a lead-lag networks is not restricted to similar states, but may happen between any two states: for example, the pair $(+1\to-1)$  is considered since the buying activity of an agent may predict the selling activity of other agents (or himself). Note that reversing the time arrow leads to the same lead-lag networks, albeit with inverted link direction.

\subsection{Lead-Lag SVNs between two timescales}
The main methodological contribution of our work is to introduce a generic framework to infer lead-lag relationships between two different timescales.

For a given time $t$, causality relationships are necessarily between actions up to $t$ and from $t$ onwards. When using a single timescale $\Delta t$, it is natural to take $t=k\Delta t$ and to check if some time series in time slice $[(k-1)\Delta t,k\Delta t[$ leads on some other time series in time slice $[k\Delta t,(k+1)\Delta t[$, and to perform this check for all possible values of $k$. This easily generalises to two consecutive intervals defined at two different timescales $\Delta_1$  and $\Delta_2$ : for a given time $t$, lead-lag may be between time slice $[t-\Delta t_1,t[$ and $[t,t+\Delta t_2[$. In the following, by convention, $\Delta t_1$ is the timescale over which agent states are determined in the past and $\Delta t_2$ the timescale in the future.

\begin{figure}
\begin{subfigure}[b]{1\textwidth}
\centering
\begin{tikzpicture}
\draw[black, thick, ->] (0,3) -- (9.5,3);
\node [right] at (9.5,3) {t};
\draw[black, thick, ->] (0,1) -- (9.5,1);
\node [right] at (9.5,1) {t};
\draw[decoration={brace,mirror,raise=5pt},decorate]
  (0,2.75) -- node[below=6pt] {$\Delta t_1$} (3,2.75);
\draw[decoration={brace,mirror,raise=5pt},decorate]
  (3,0.75) -- node[below=6pt] {$\Delta t_2$} (5,0.75);
\draw[black, thick] (0,2.8) -- (0,3.2);
\draw[black, thick] (3,2.8) -- (3,3.2);
\draw[black, thick] (6,2.8) -- (6,3.2);
\draw[black, thick] (9,2.8) -- (9,3.2);
\draw[black, thick] (0,0.8) -- (0,1.2);
\draw[black, thick] (3,0.8) -- (3,1.2);
\draw[black, thick] (6,0.8) -- (6,1.2);
\draw[black, thick] (9,0.8) -- (9,1.2);
\draw[black, thick] (1.5,3) ellipse (1.5 and 0.25);
\draw[black, thick] (4.5,3) ellipse (1.5 and 0.25);
\draw[black, thick] (7.5,3) ellipse (1.5 and 0.25);
\draw[black, thick] (4,1) ellipse (1 and 0.25);
\draw[black, thick] (7,1) ellipse (1 and 0.25);
\draw[black, thick, ->] (1.5,2.70) -- (4,1.30);
\draw[black, thick, ->] (4.5,2.70) -- (7,1.30);
\end{tikzpicture}
\caption{}\vspace{1cm}
\end{subfigure}
\begin{subfigure}[b]{1\textwidth}
\centering
\begin{tikzpicture}
\draw[black, thick, ->] (0,3) -- (9.5,3);
\node [right] at (9.5,3) {t};
\draw[black, thick, ->] (0,1) -- (9.5,1);
\node [right] at (9.5,1) {t};
\draw[decoration={brace,mirror,raise=5pt},decorate]
  (1,2.75) -- node[below=10pt] {$\Delta t_1$} (3,2.75);
\draw[decoration={brace,mirror,raise=5pt},decorate]
  (3,0.75) -- node[below=6pt] {$\Delta t_2$} (6,0.75);
\draw[black, thick] (0,2.8) -- (0,3.2);
\draw[black, thick] (3,2.8) -- (3,3.2);
\draw[black, thick] (6,2.8) -- (6,3.2);
\draw[black, thick] (9,2.8) -- (9,3.2);
\draw[black, thick] (0,0.8) -- (0,1.2);
\draw[black, thick] (3,0.8) -- (3,1.2);
\draw[black, thick] (6,0.8) -- (6,1.2);
\draw[black, thick] (9,0.8) -- (9,1.2);
\draw[black, thick] (2,3) ellipse (1.0 and 0.25);
\draw[black, thick] (5,3) ellipse (1.0 and 0.25);
\draw[black, thick] (8,3) ellipse (1.0 and 0.25);
\draw[black, thick] (4.5,1) ellipse (1.5 and 0.25);
\draw[black, thick] (7.5,1) ellipse (1.5 and 0.25);
\draw[black, thick, ->] (1.5,2.70) -- (4,1.30);
\draw[black, thick, ->] (4.5,2.70) -- (7,1.30);
\end{tikzpicture}
\caption{}\vspace{1cm}
\end{subfigure}
\caption{Schematic diagram showing how lead-lag links are established when (a) $\Delta t_1>\Delta t_2$ and (b) $\Delta t_1<\Delta t_2$. In both cases, the alignment is on $\Delta t_M=\max(\Delta t_1,\Delta t_2)$.}
\label{fig:segmentation_leadlag}
\end{figure}
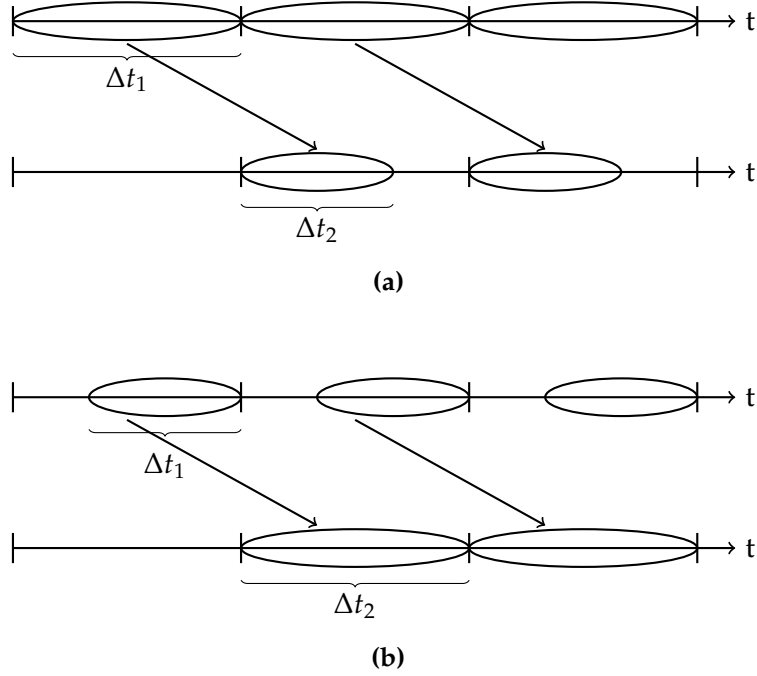

Once the states of the agents are determined in both intervals, $t$ is then shifted. The only way to avoid using overlapping time intervals is to shift time by $\Delta t_M=\max(\Delta t_1,\Delta t_2)$ (see Fig.\ \ref{fig:segmentation_leadlag}), hence $t=k\Delta_M$.

Grouping the agents, i.e., labelling them, is performed for each timescale separately: one first applies SVNs with $\Delta t_1$, which yields the groups denoted by $G_1$ and then with $\Delta t_2$, which yields $G_2$.  Finally, the state of $\sigma^{(1)}_{g_1}(k\Delta t_M)$ of each group $g_1\in G_1$ is computed in time slice $[k\Delta t_M-\Delta t_1,k\Delta t_M[$ for all the values of $k$ and the state $\sigma^{(2)}_{g_2}(k\Delta t_M)$ of each group $g_2\in G_2$ is computed in time slice $[k\Delta t_M,k\Delta t_M+\Delta t_2[$. 

Finally, lead-lag SVNs are obtained for each chosen state pair $(\sigma_1,\sigma_2)$ according to the coincidence of states of $(\sigma^{(1)}_{g_1}(k\Delta t_M)=\sigma_1,\sigma^{(2)}_{g_2}(k\Delta t_M)=\sigma_2)$.

\section{Data sets}
Our data sets contain trader-resolved transactions of the EUR/USD currency pair and come from two independent sources: Swissquote Bank SA (SQ hereafter), a Swiss broker-dealer with a large market share in foreign exchange (FX) transactions in Switzerland, and a large anonymous dealer bank which serves major institutional clients (LB hereafter).
Both data sets list all the trades of their clients: traded currency pair, anonymous client identification number, trade time (at a millisecond resolution), signed volume, and the FX transaction rate. 
We focus on the EUR/USD pair as it is one of the most traded pairs in both data sets.
A summary of the data sets structure and contents is provided
in Table \ref{table:statistics_data sets}.

\begin{table}[ht]
\begin{center}
    \begin{tabular}{ | l | l | l | p{1.5cm} |}
    \hline
    \textbf{data set} &  \textbf{Timespan} & \textbf{Traders} & \textbf{Trades} \\ \hline
    LB & 01 Jan. 2013 - 15 Sep. 2014 & $>10^3$ & $>10^5$ \\ \hline
    SQ & 01 Jan. 2014 - 15 Jun. 2016 & $>10^3$ & $>10^7$\\ \hline
    \end{tabular}
    \caption[Table caption text]{Data sets for EUR/USD currency pair.}
\label{table:statistics_data sets}
\end{center}
\end{table}

While FX markets never close, transactions are  rare during nights and week-ends. We thus focus on active hours, i.e., from 9:00 to 17:00 on week days. We only look for links between adjacent timeslices on the same day in order to avoid spurious boundary effects or overnight lead-lag links.


\section{Results}

Because the activity of each trader is not regular in either data set, one cannot use the whole data sets to infer lead-lag networks. We use here rolling calibration time windows of $T_{\textrm{in}}=\{30,60,90,120\}$ business days for LB and $T_{\textrm{in}}=\{30,60\}$ business days for SQ because computations on this much larger dataset are expensive (several days for a single value of $T_{\textrm{in}}$).
For each time window, we take each pair of timescales $\Delta t_1$ and $\Delta t_2$ belonging to the arithmetic sequence from 300 seconds (5 minutes) to 14400 seconds (4 hours) with a step of 5 minutes (which corresponds to 1176 unique pairs of timescales). 
We use seconds as the unit of time in the figures. 

\subsection{$\Delta t_1=\Delta t_2$}

\begin{figure}[ht]
\centering
\begin{subfigure}{.49\textwidth}
\centering
\caption{LB}
\includegraphics[scale=0.25]{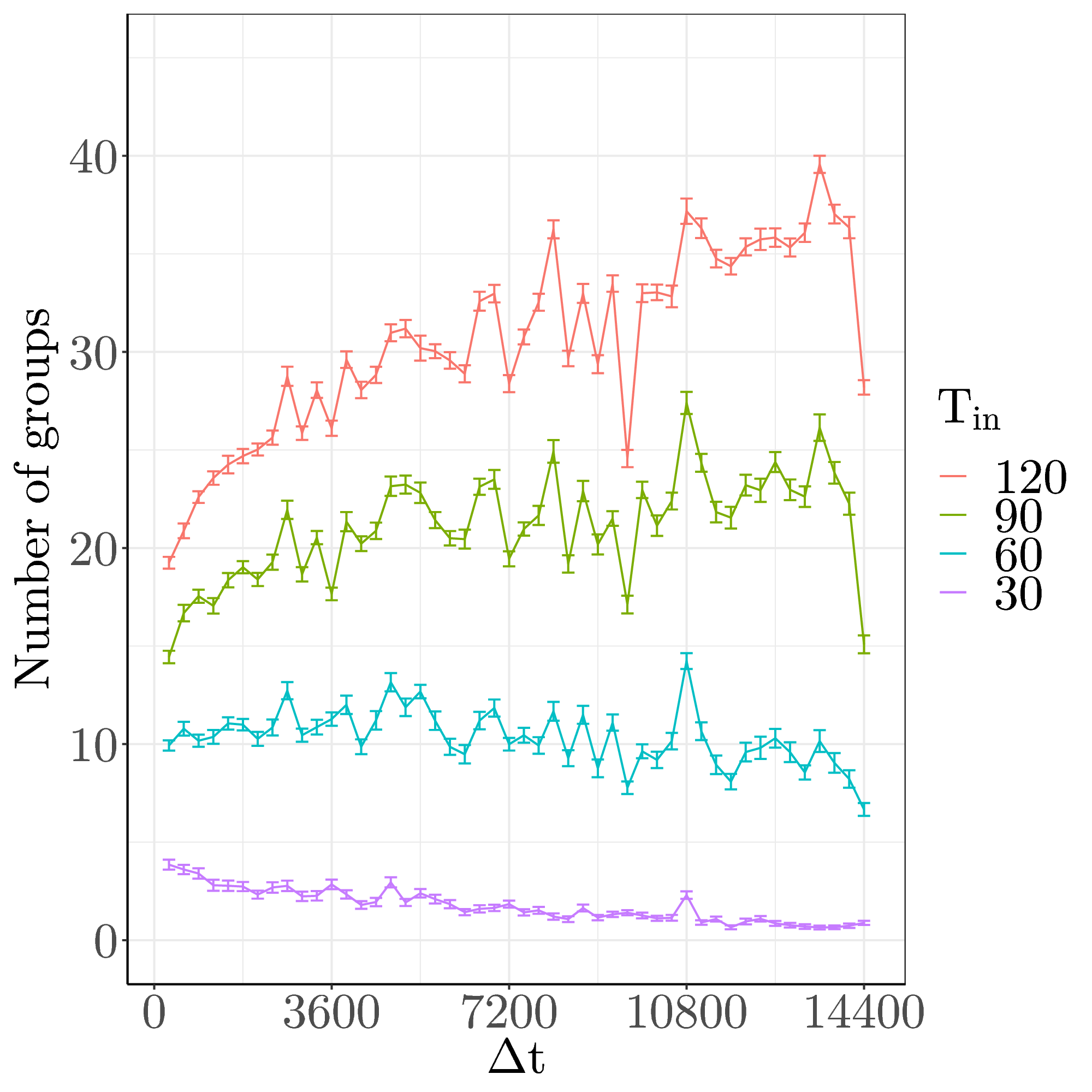}
\end{subfigure}
\begin{subfigure}{.49\textwidth}
\centering
\caption{SQ}
\includegraphics[scale=0.25]{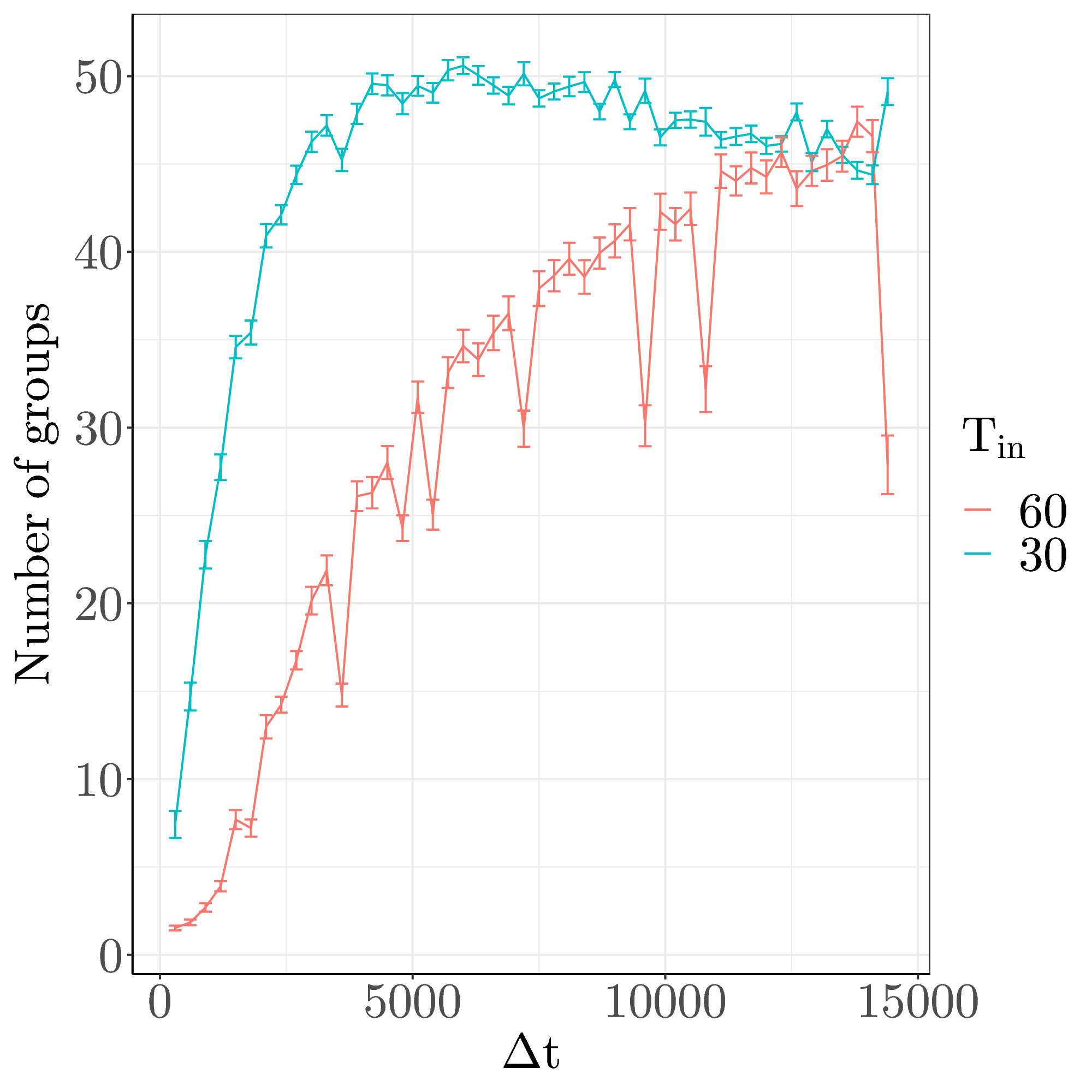}
\end{subfigure}
\caption{Average number of groups as a function of coarsening timescale $\Delta t$ seconds and the number of days in the calibration window $T_{\textrm{in}}$.
$\Delta t = \Delta t_1=\Delta t_2$.}
\label{fig:groups_sizes}
\end{figure}

\begin{figure}[ht]
\centering
\begin{subfigure}{.49\textwidth}
\centering
\caption{LB}
\includegraphics[scale=0.25]{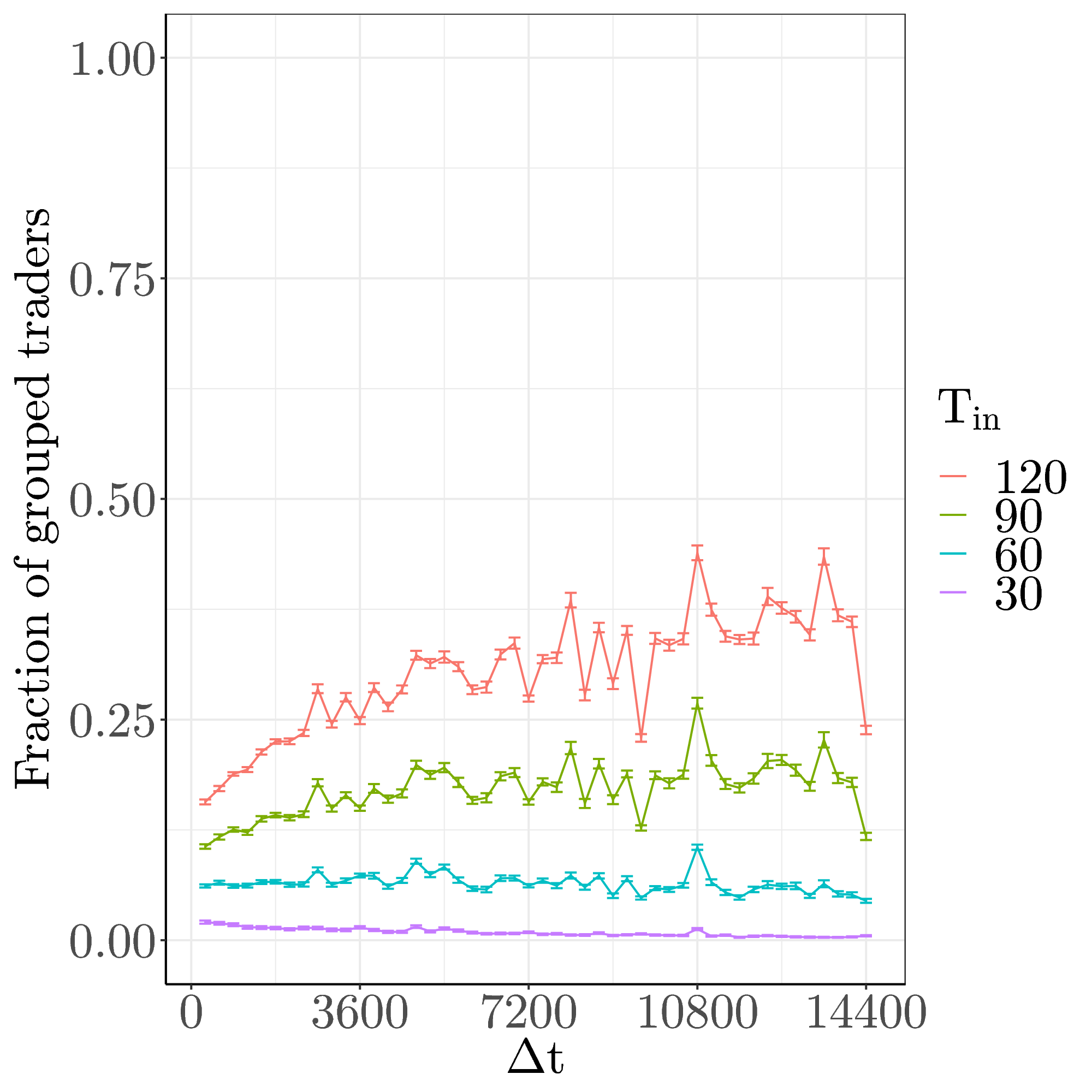}
\end{subfigure}
\begin{subfigure}{.49\textwidth}
\centering
\caption{SQ}
\includegraphics[scale=0.25]{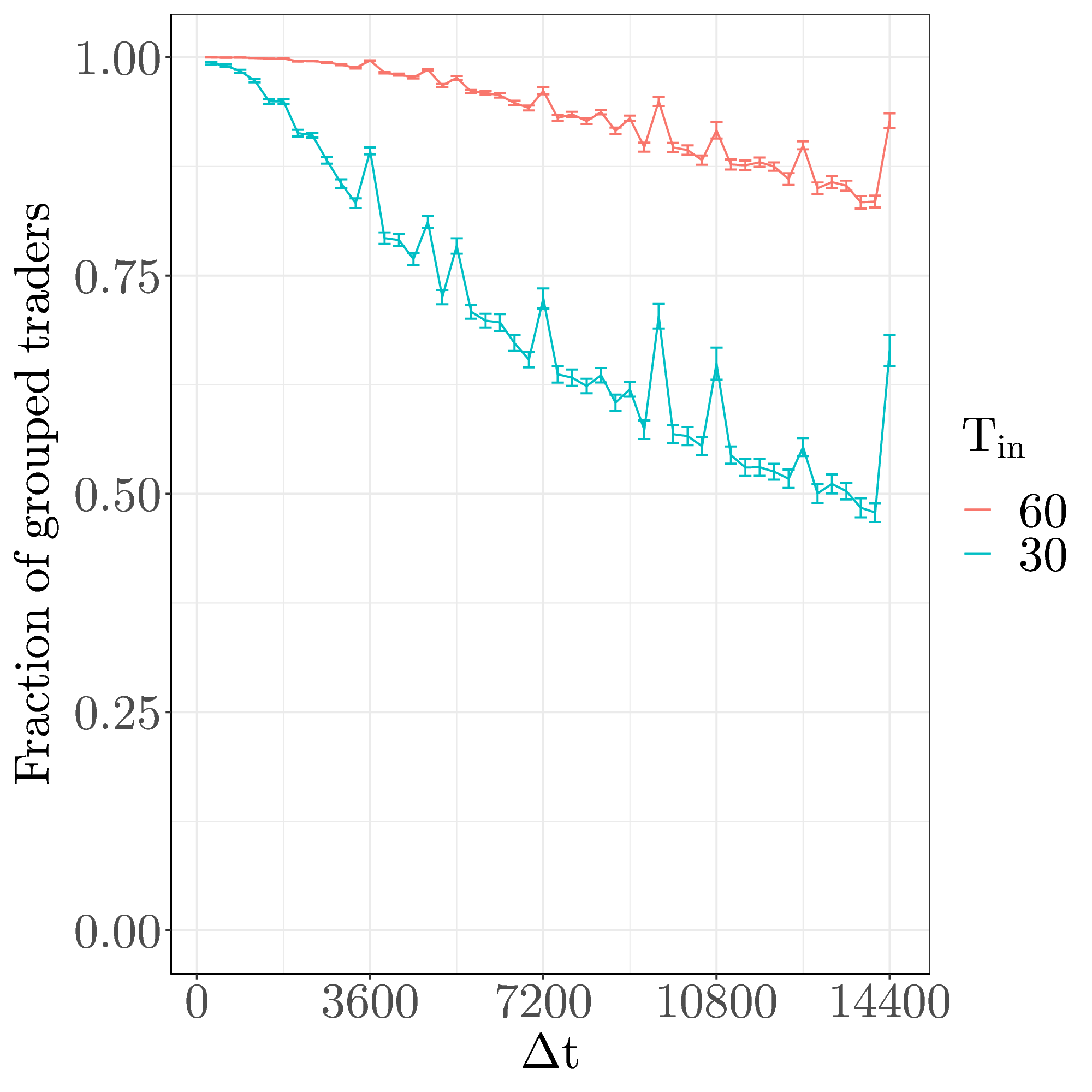}
\end{subfigure}
\caption{Average fraction of traders grouped by SVNs as a function of $\Delta t$ and $T_{\textrm{in}}$. 
$\Delta t = \Delta t_1=\Delta t_2$.}
\label{fig:fraction_active_traders}
\end{figure}

\begin{figure}[ht]
\centering
\begin{subfigure}{.49\textwidth}
\centering
\caption{LB}
\includegraphics[scale=0.25]{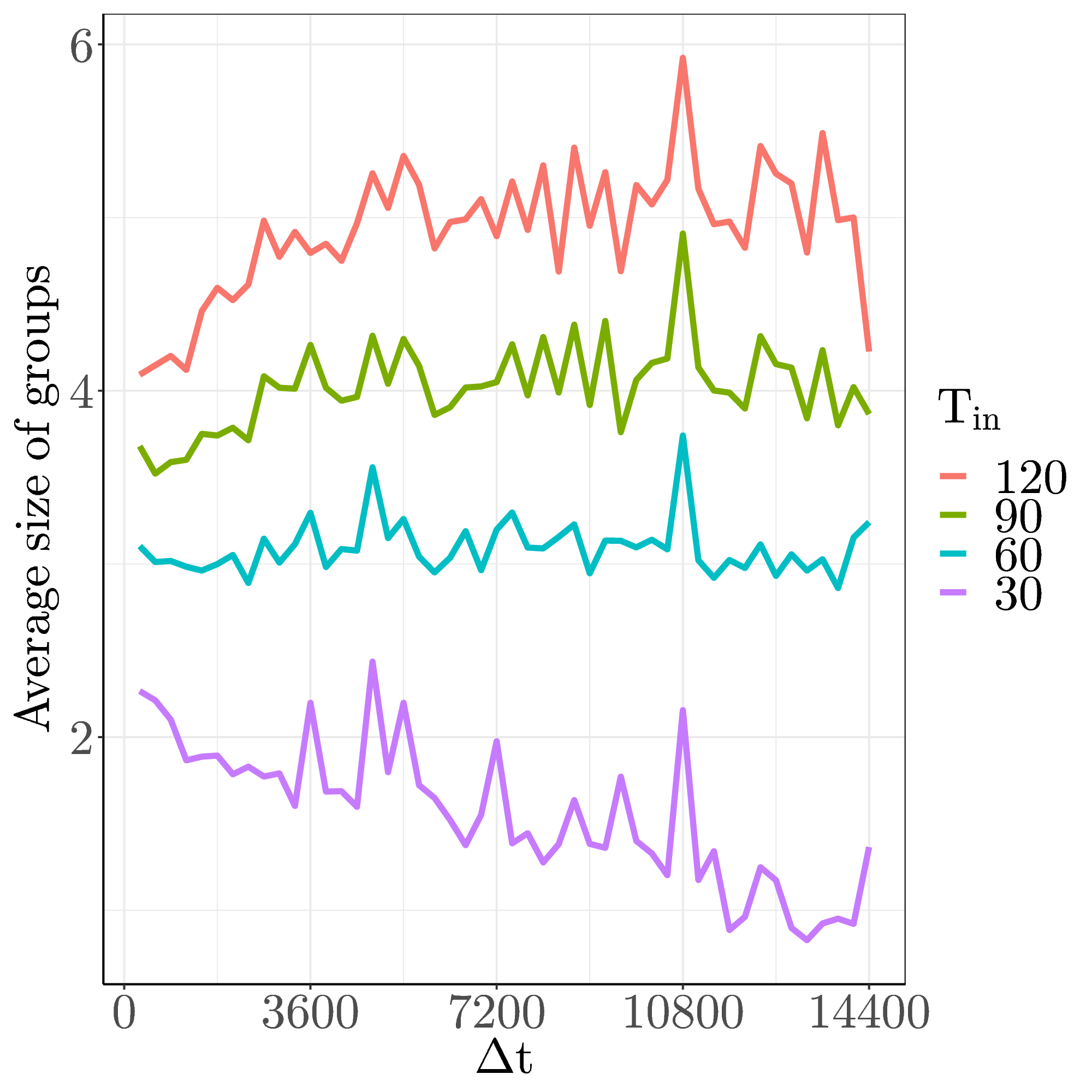}
\end{subfigure}
\begin{subfigure}{.49\textwidth}
\centering
\caption{SQ}
\includegraphics[scale=0.25]{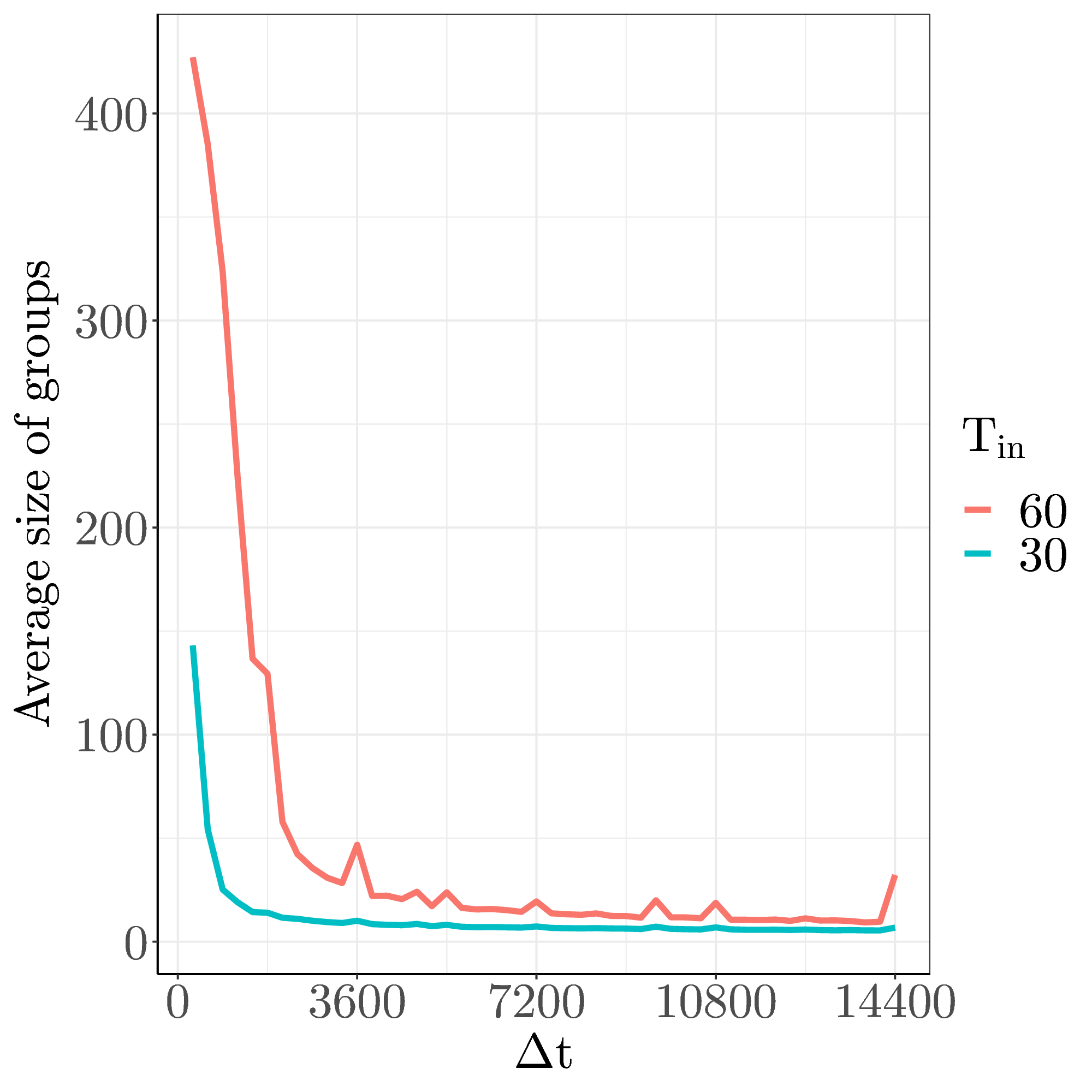}
\end{subfigure}
\caption{Average size of groups as a function of time coarsening $\Delta t$ (seconds) and calibration window length $T_{\textrm{in}}$ (days).
$\Delta t = \Delta t_1=\Delta t_2$ .}
\label{fig:avg_size_groups}
\end{figure}

A systematic study of the $\dtone=\dttwo=\Delta t<$1 day  case is missing in the literature mainly because intraday data sets are not readily available and this part fills this gap. As it turns out, the dependence of lead-lag networks on $\Delta t$ and $T_{\textrm{in}}$ is notably different between retail and institutional clients in our data sets.

Figure \ref{fig:groups_sizes} plots the number of groups averaged over all the calibration windows as a function of $\Delta t$ for all $T_{\textrm{in}}$, for LB and SQ. 
The number of groups found by the LL-SVNs and InfoMap is a measure of the statistically validated diversity of behavior and of the potential richness of connectivity. For example, only a few groups of LB clients for $T_{\textrm{in}}=30$ days and large $\Delta t$ are detected, while the largest value of $T_{\textrm{in}}=120$ days yields the most groups for LB: the larger $T_{\textrm{in}}$, the larger the number of groups in this data set.
 One also sees a sudden drop of the number of groups for $\Delta t=14400\textrm{ seconds}=4\textrm{ hours}$, which is likely a by-product of the fact that we keep 8 hours of trading each day.
The number of groups of SQ retail clients behaves in the  opposite way: the smaller $T_{\textrm{in}}$, the larger the number of groups. 

Figure \ref{fig:fraction_active_traders} once again displays a striking difference between SQ and LB clients. While most SQ clients belong in a group, the fraction of LB clients that belong in a cluster is much smaller and increases with $T_{\textrm{in}}$. A possible explanation is that the percentage of SQ clients that use similar algorithmic trading strategies is much higher, which makes it easier to group them. 

The group size distribution is very skewed, particularly for SQ traders: for example the median size of the groups is much smaller that the average group size, which is due to the fact that most groups contain 2-5 traders, while a few groups contain many traders, especially for small $\Delta t$ (see Fig.\ \ref{fig:avg_size_groups}). The disparition of this large group for larger $\Delta t$ is likely due to the fact that the typical time between two trades is smaller than 1 hour for SQ traders: summarising the activity of traders with only a few states at timescales larger than 1 hour  causes a notable loss of descriptive power.

The {\em raison d'être} of calibration with sliding windows is  the non-stationarity not only of the population of traders, but also of their behavior. 
If both are roughly stationary, a longer $T_{\textrm{in}}$, at fixed $\Delta t$, should give more precise and richer results, and inversely.
This is likely a major cause of the difference between SQ and LB traders, the latter behaving in a much more stationary manner.

\begin{figure}
\centering
\begin{subfigure}{.49\textwidth}
\centering
\caption{LB, $T_{\textrm{in}}=60$ days}
\includegraphics[scale=0.25]{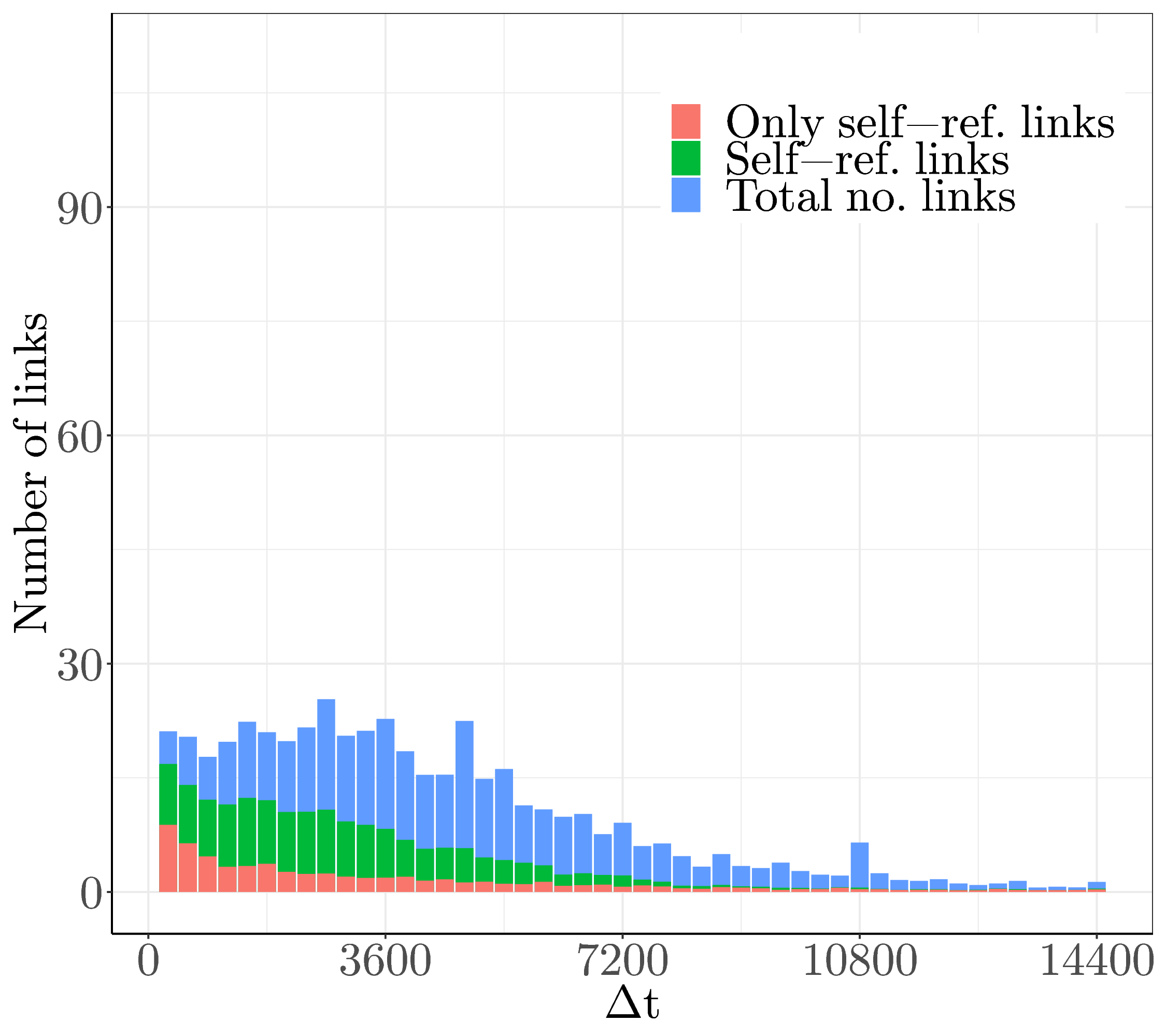}
\end{subfigure}
\begin{subfigure}{.49\textwidth}
\centering
\caption{LB, $T_{\textrm{in}}=120$ days}
\includegraphics[scale=0.25]{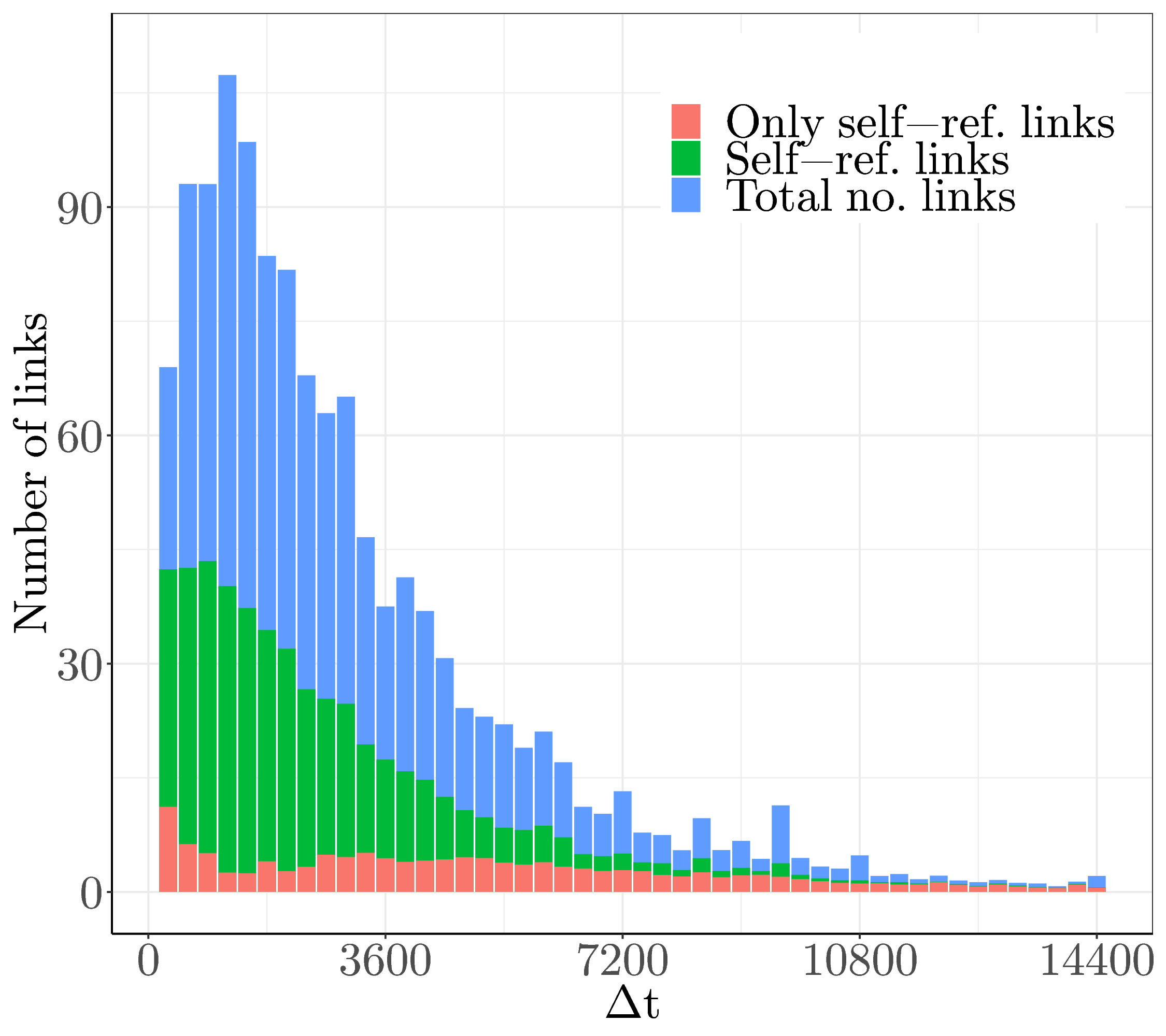}
\end{subfigure}
\vskip\baselineskip
\begin{subfigure}{.49\textwidth}
\centering
\caption{SQ, $T_{\textrm{in}}=30$ days}
\includegraphics[scale=0.25]{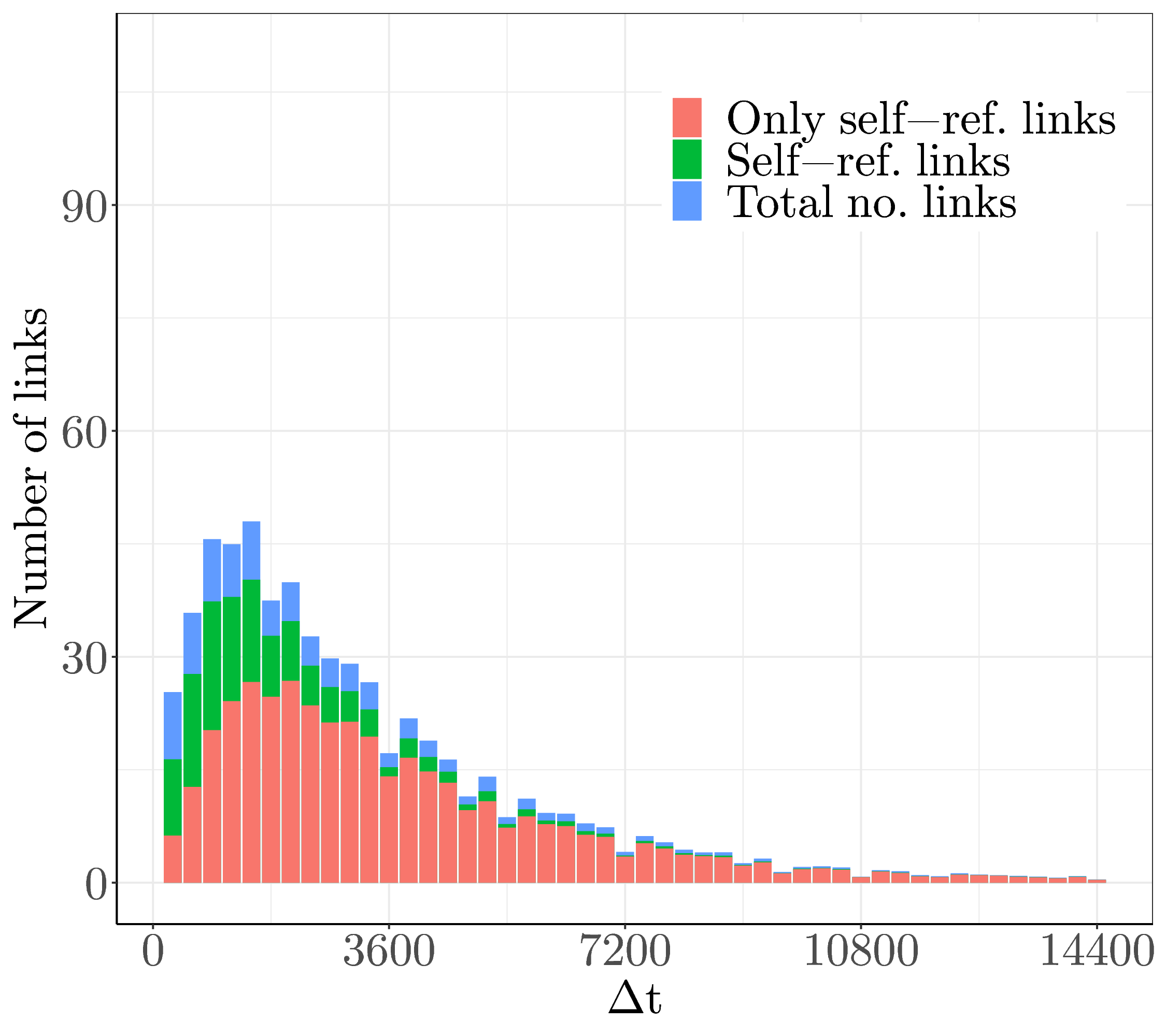}
\end{subfigure}
\begin{subfigure}{.49\textwidth}
\centering
\end{subfigure}
\begin{subfigure}{.49\textwidth}
\centering
\caption{SQ, $T_{\textrm{in}}=60$ days}
\includegraphics[scale=0.25]{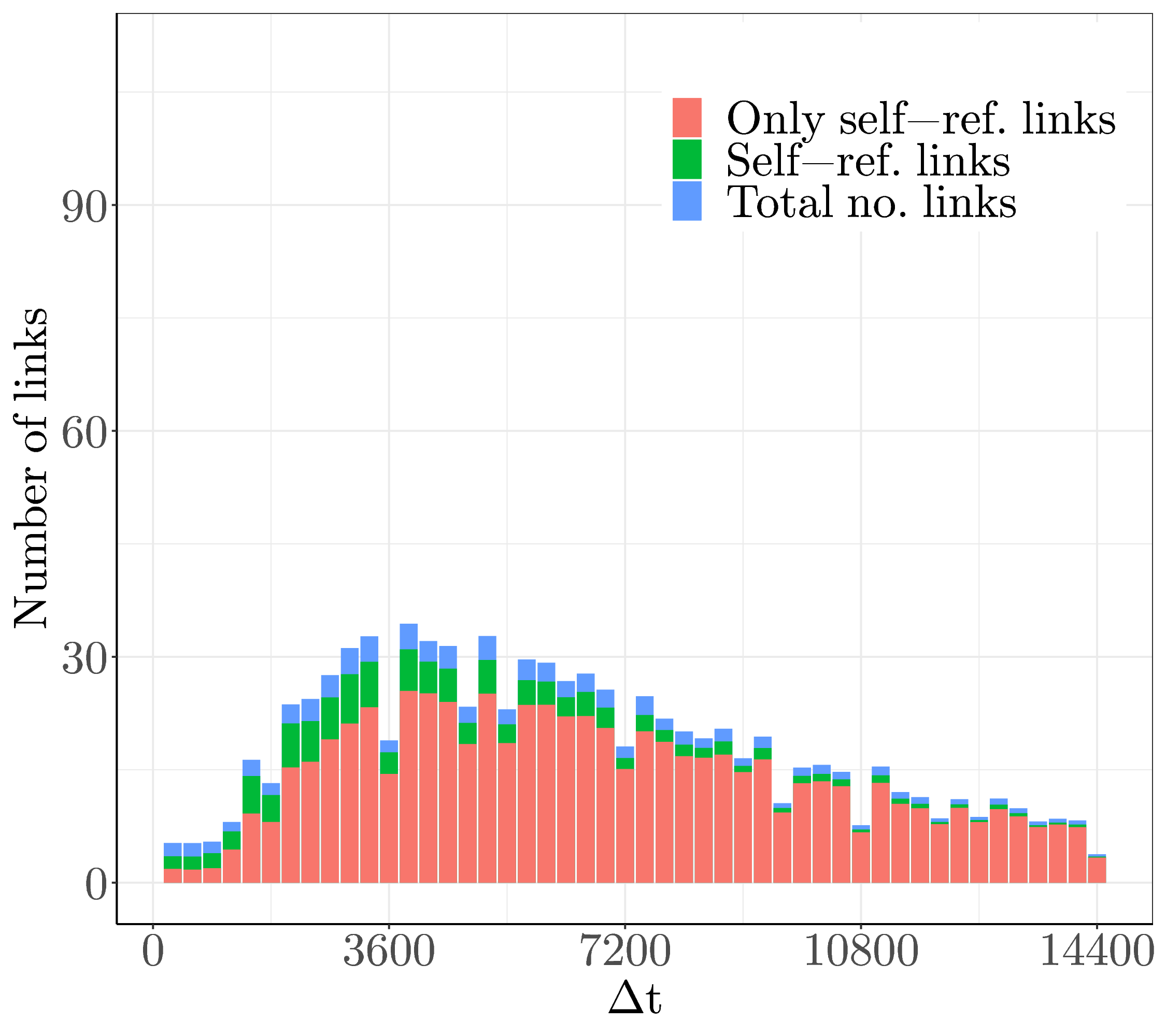}
\end{subfigure}
\caption{Total number of links as a function of the timeslice duration (in seconds) $\Delta t=\dtone=\dttwo$, for groups with only self-referential links, self-referential links and links to other groups and only links to other groups.
\label{fig:self_ref}}
\end{figure}

Let us now turn to the links themselves. Since we deal with lead-lag networks, they are directed. Links can be of two types: either from one group to another one, or to the same group, which we call a self-referential links. Occasionally, some groups only link to themselves, which would happen if they use a strategy whose activity does not systematically lead on another one, but whose activity, on average, occurs at a scale comparable to $\Delta t$.
  
Figure \ref{fig:self_ref} reports the average total number of lead-lag links  both LB and SQ. Some groups only have self-referential links (which are labelled as 'only self-referential links' by a slight abuse of language), some others have both types of links. The lead-lag networks of the two types of traders are clearly different: the typical fraction of groups with only self links is small for LB traders, but much larger for SQ traders.
The timeslice length $\Delta t$ influences the number of non-self-referential links for both populations: their number decreases when $\Delta t >\textrm{1 hour}$ and are negligible at resolutions coarser than 10800 seconds (three hours) for LB data. In LB data, lead-lag networks contain a maximum of links for $T_{\textrm{in}}=120$ and $\Delta t\simeq 1800s$ (30 minutes). For SQ data, the choice  $T_{\textrm{in}}=60$ has the interesting property that the number of links is spread over a wider range of $\Delta t$ than for $T_{\textrm{in}}=30$.

Given the results above, we will thus focus on $T_{\textrm{in}}=120$ days for LB and $T_{\textrm{in}}=60$ days for SQ as these choices yield the most interesting networks (Figs \ref{fig:groups_sizes} and \ref{fig:self_ref}), while including a sizeable fraction of clients in clusters (Fig. \ref{fig:fraction_active_traders}).

\subsection{$\Delta t_1\neq \Delta t_2$}

\subsubsection{Links}

When $\Delta t_1\ne \Delta t_2$, both timescales may influence each other in an asymmetric way. 
Our strategy is to capture such an asymmetry by using several quantities related to both the directed network structure and the rate of trading. Each quantity is estimated for each pair $(\Delta t_1,\Delta t_2)$, each of them ranging from 5 minutes (300 seconds) to 4 hours (14440 seconds) by steps of 5 minutes, which gives 1176 unique pairs.
Since we measure these quantities over many calibration windows, we obtain a timeseries for each quantity and for each pair.

\begin{figure}
\centering

\begin{subfigure}{.49\textwidth}
\centering
\caption{Number of links}
\includegraphics[scale=0.35]{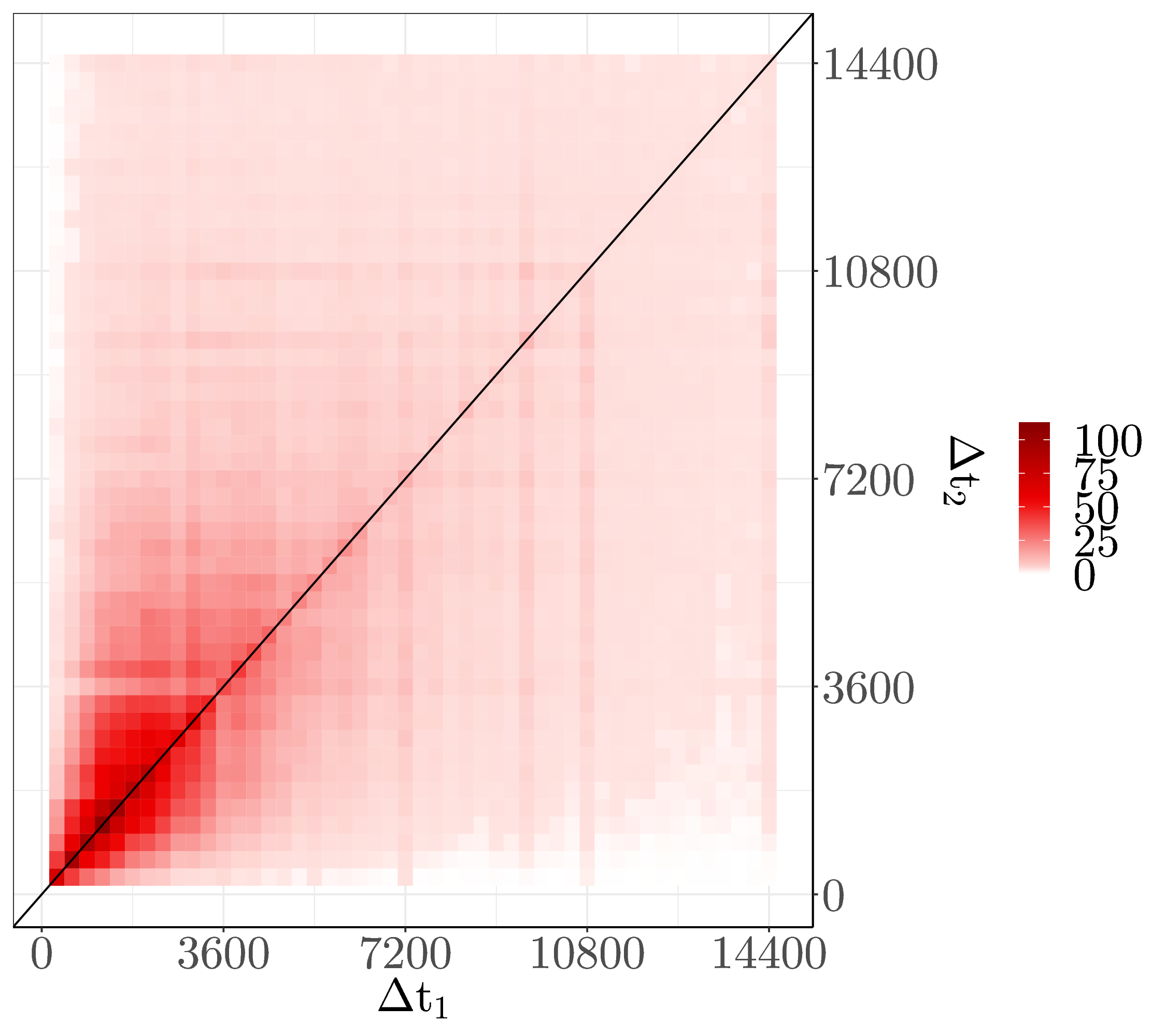}
\end{subfigure}
\begin{subfigure}{.49\textwidth}
\centering
\caption{FDR-corrected t-stat}
\includegraphics[scale=0.35]{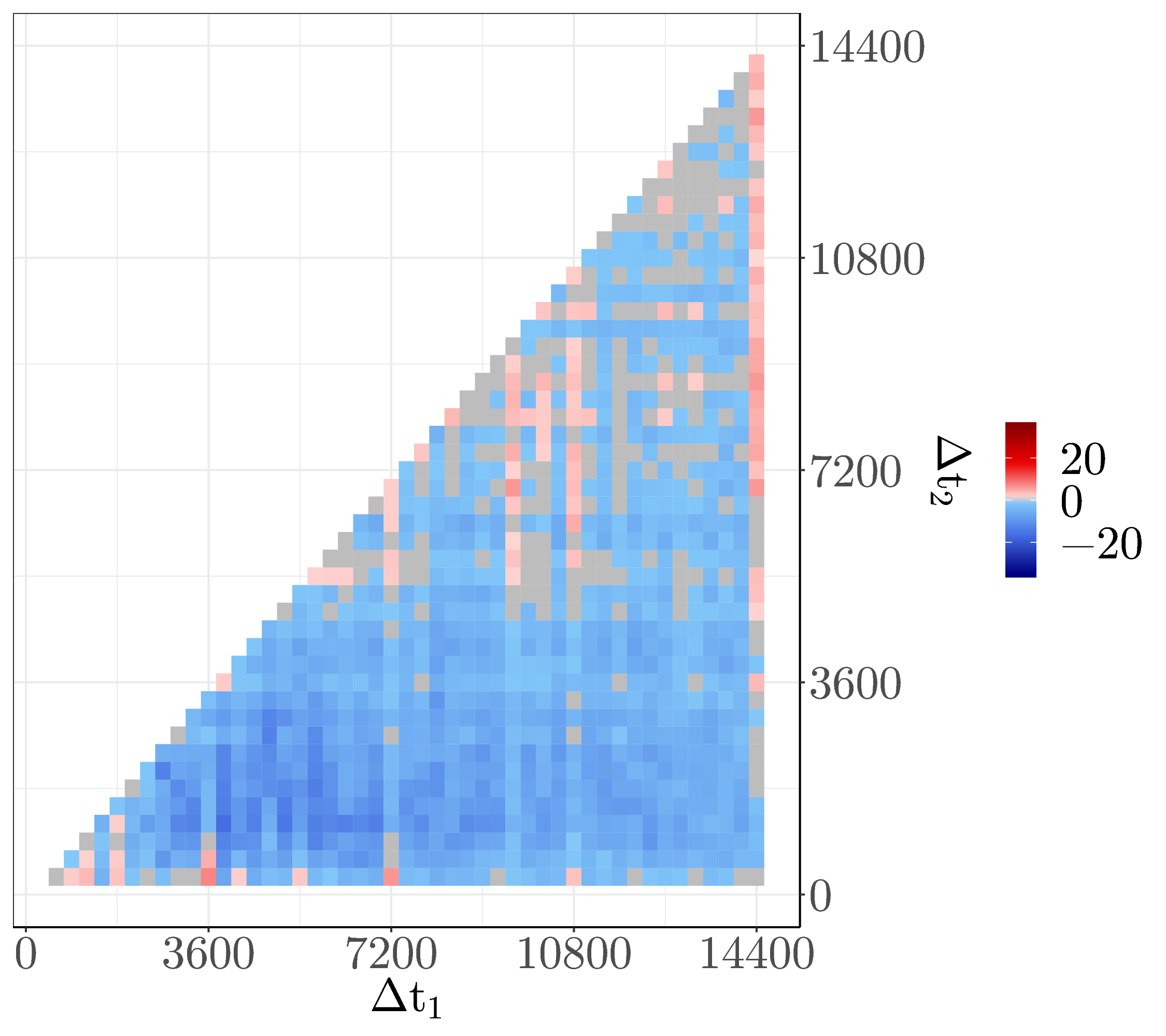}
\end{subfigure}

\caption{Left hand-side plot: average number of lead-lag links for LB ($\dtone$ leads on $\dttwo$). Right hand-side plot: t-statistics of the difference between the number of links of the pairs $(\Delta t_1,\Delta t_2)$ and  $(\Delta t_2,\Delta t_1)$; negative values indicate that shorter timescales link significantly more to longer timescales.
$T_{\textrm{in}}=120$ days}
\label{fig:mean_links_LB}
\end{figure}

\begin{figure}
\centering
\begin{subfigure}{.49\textwidth}
\centering
\caption{Number of links}
\includegraphics[scale=0.35]{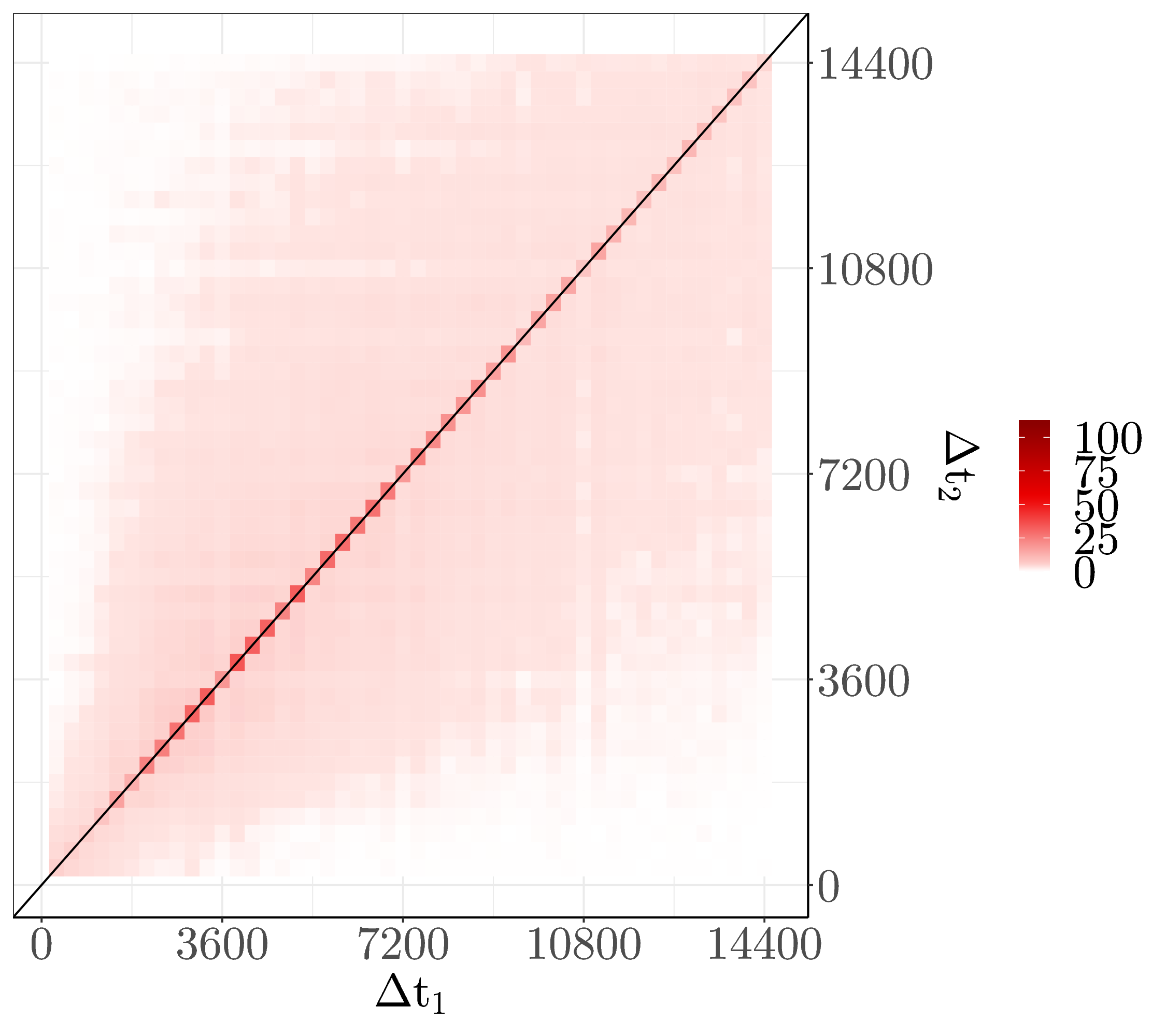}
\end{subfigure}
\begin{subfigure}{.49\textwidth}
\centering
\caption{FDR-corrected t-stat}
\includegraphics[scale=0.35]{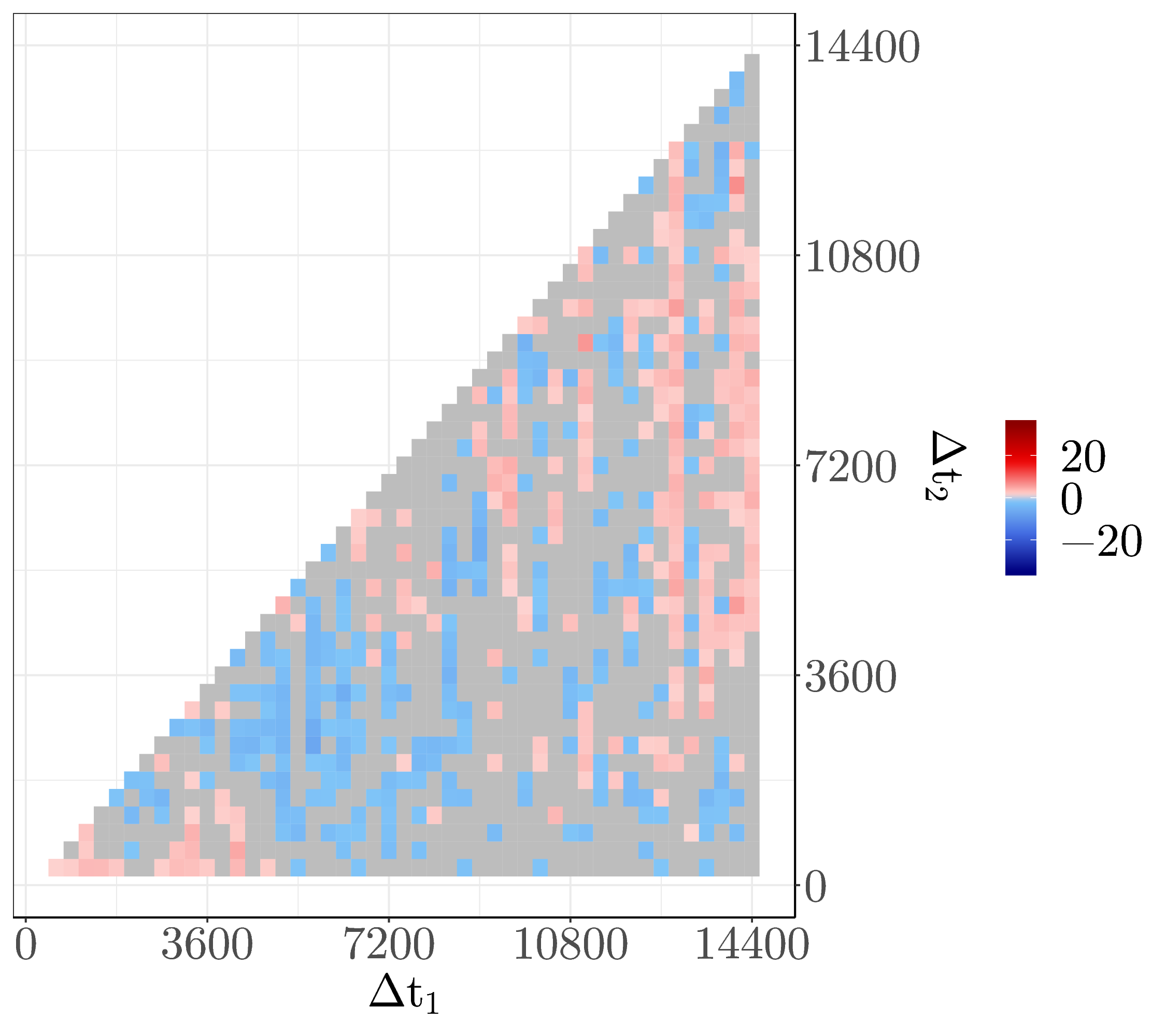}
\end{subfigure}
\vskip\baselineskip
\caption{Left hand-side plot: average number of lead-lag links for SQ ($\dtone$ leads on $\dttwo$). Right hand-side plot: t-statistics of the difference between the number of links of the pairs $(\Delta t_1,\Delta t_2)$ and  $(\Delta t_2,\Delta t_1)$; negative values indicate that shorter timescales link significantly more to longer timescales. Gray areas correspond to non-significant relationship according to a FDR of $0.05$.
$T_{\textrm{in}}=60$ days.}
\label{fig:mean_links_SQ}
\end{figure}

Let us first start with the number of links.
The left hand side plots of Figs.~\ref{fig:mean_links_LB} and \ref{fig:mean_links_SQ} show the average number of links for each pair of timescales.
Let us remind the convention regarding the labelling of timescales: $\Delta t_1$ (on the x-axis) leads on $\Delta t_2$ (on the y-axis): as a consequence, points above the $\Delta t_1=\Delta t_2$ diagonal correspond to smaller timescales leading on longer timescales, and inversely.
It is useful to keep in mind that on that diagonal the number of links is maximal for small values of $\Delta t$ for both LB and SQ (Fig.  \ref{fig:self_ref}).

In accordance with Fig.~\ref{fig:self_ref}, there are more links for smaller values of $\Delta t_1$ and $\Delta t_2$ around the diagonal. There are also more links for some particular values of either $\dtone$ or $\dttwo$, e.g. multiples of full hours. This may indicate that some traders have a typical activity change over 1 hour, e.g. a trading strategy that depends on the time of the day, or that they trade between, say, 9:00 and 10:00, 10:00 and 11:00, and so on, or that the maximum holding period of some strategies is one, two, or three hours.

At least for LB, the red area around $(3000,3800)$ in Fig.\ref{fig:links_micro_cor_LB}a is significantly larger and darker than the one around $(3000,3800)$, which implies that there are on average more links from shorter timescales to longer timescales. 
The statistical significance of this difference is assessed in the following way: let us denote the number of links of the pair $(\Delta t_1,\Delta t_2)$, the first timescale of the pair leading on the second one, in calibration window $i$ by $W_i(\Delta t_1,\Delta t_2)$. One  applies a t-statistics to the timeseries of the difference $\delta W_i(\Delta t_1,\Delta t_2)=W_i(\Delta t_1,\Delta t_2)-W_i(\Delta t_2,\Delta t_1)$. In order to avoid too many false positives, we use an FDR correction of the threshold p-value for multiple hypotheses made in this plot (and in all similar plots), setting the rate at FDR$=0.05$ (see subsection \ref{sec:SVN} for FDR in relationship with multiple hypothesis testing). 
The right plots of Figs.\ \ref{fig:mean_links_LB} and \ref{fig:mean_links_SQ} display the selected t-stats of $\delta W_i(\Delta t_1,\Delta t_2$) (non-grey points): blue zones correspond to lead-lag links from shorter to longer timescales, and reversely for red zones.

The plots for LB are overwhelmingly blue: there are more links from short timescales to long timescales. There is a clear exception for $\Delta t_1=14400$ seconds (4 hours), which once again is probably a by-product of keeping exactly 8 hours of data each day. 
One notes however small red regions at around $(10800, 7200)$,  $(3600, 300)$ and $(7200, 300)$. For the SQ traders, the link structure is sparser and more complex, while clearly significantly. 

The number of links themselves are not sufficient to characterize the lead-lag between timescales for traders. 
Indeed, it may happen that a group has more than one link to another group, even for the same initial state. 
For example, group 1 may have links $1\to 1$ and $1\to-1$ to group 2. The presence of such dual links means in this case that the mostly buying activity of group 1 triggers either state $+1$ or $-1$ of group 2. 
In other words, it triggers a directional activity of group 2, whose sign is indeterminate from the knowledge of the state of group 1.  
In a prediction setting, dual links reduce the prediction power, but as long as enough single links do exist, order flow prediction is possible, as shown by \citet{challet2018trader}. 

\subsubsection{From activity TRA to price volatility TRA}

Time reversal asymmetry of prices, while being totally intuitive, is not totally trivial to measure owing to the amount of noise in financial data. \citet{zumbach2001heterogeneous} proposed to measure the TRA between historical volatility measured over $\Delta t_1$  and realized volatility, estimated over $\Delta t_2$. More precisely, for a given $t$, one estimates the historical volatility $v_h(t)$ in the interval $]t-\Delta t_1,t[$ and the realized volatility $v_r(t)$ in the interval $[t,t+\Delta t_2[$; then one estimates the correlation of $v_r$ and $v_h$ for all chosen times $t$, denoted by $\rho_v(\Delta t_1,\Delta t_2)$. This results in volatility correlation mugshots which demonstrate the asymmetry of $\rho_v(\Delta t_1,\Delta t_2)$ with respect to the diagonal $\Delta t_1=\Delta t_2$. \cite{zumbach2009time} investigates further the TRA of volatility and proposes two more measures of TRA by noticing that the price returns in the time intervals over which volatility is estimated can be defined according to their own timescale, whose fine structure is investigated, e.g. in \citet{chicheportiche2014fine}. 
 
Our lead-lag method filters significant group activity lead-lag, not price volatility lead-lag. Therefore, connecting the two requires additional steps. First, let us remark that when time subordination holds \citep{clark1973subordination}, volatility is an increasing function of activity, all other things being equal: for example, if the volatility per trade is locally constant, then the volatility in a time interval depends on the number of trades occurring in that period of time assuming that prices are diffusive. This view is an approximation that neglects jumps of various origins, e.g. microstructural noise due to heavy-tailed distribution gaps in limit order books \citep{gillemot2006there}. However, it suggests to relate activity TRA  to volatility TRA. 

\begin{figure}
\centering

\begin{subfigure}{.49\textwidth}
\centering
\caption{Average correlation}
\includegraphics[scale=0.35]{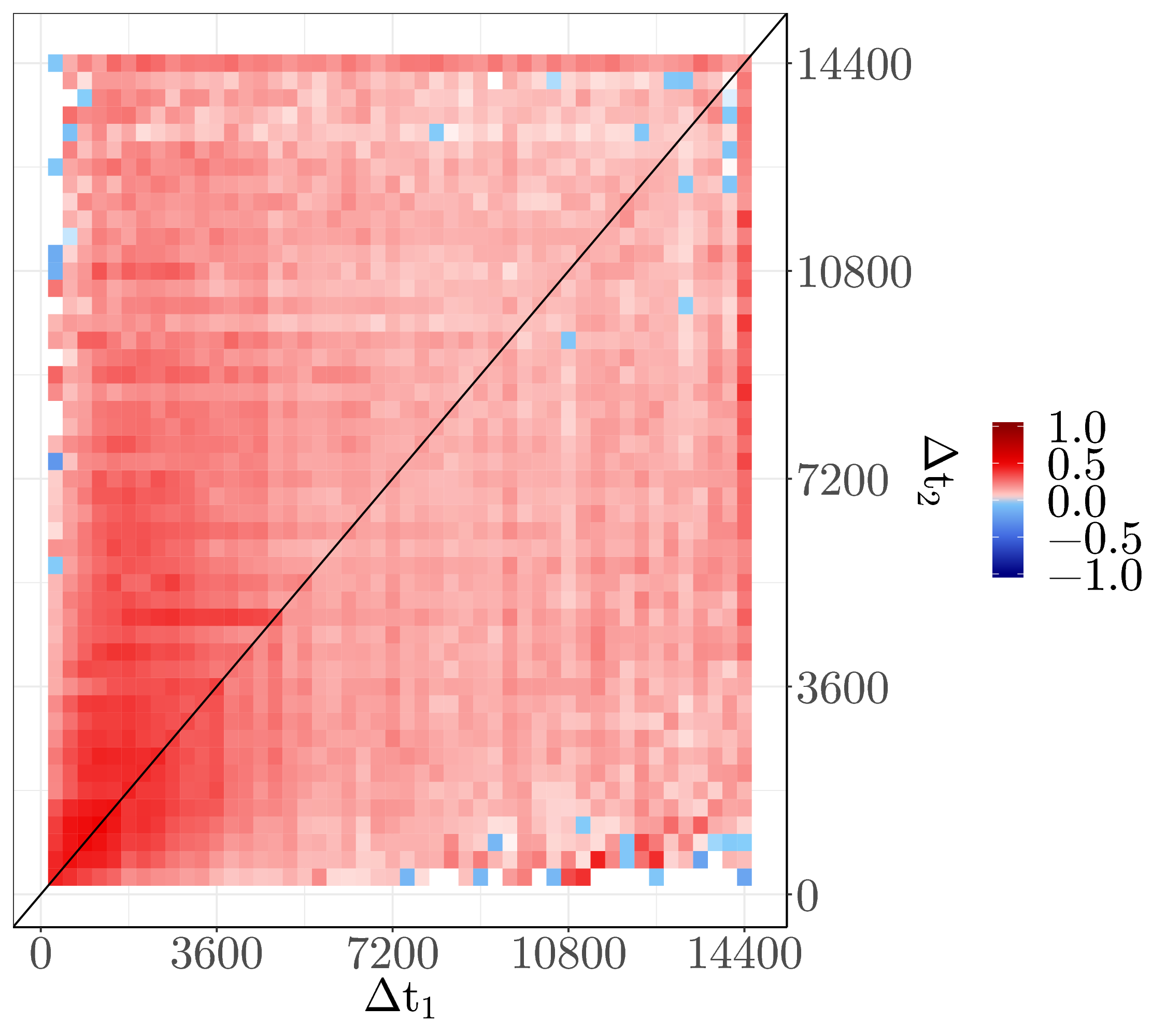}
\end{subfigure}
\begin{subfigure}{.49\textwidth}
\centering
\caption{FDR-corrected t-stat}
\includegraphics[scale=0.35]{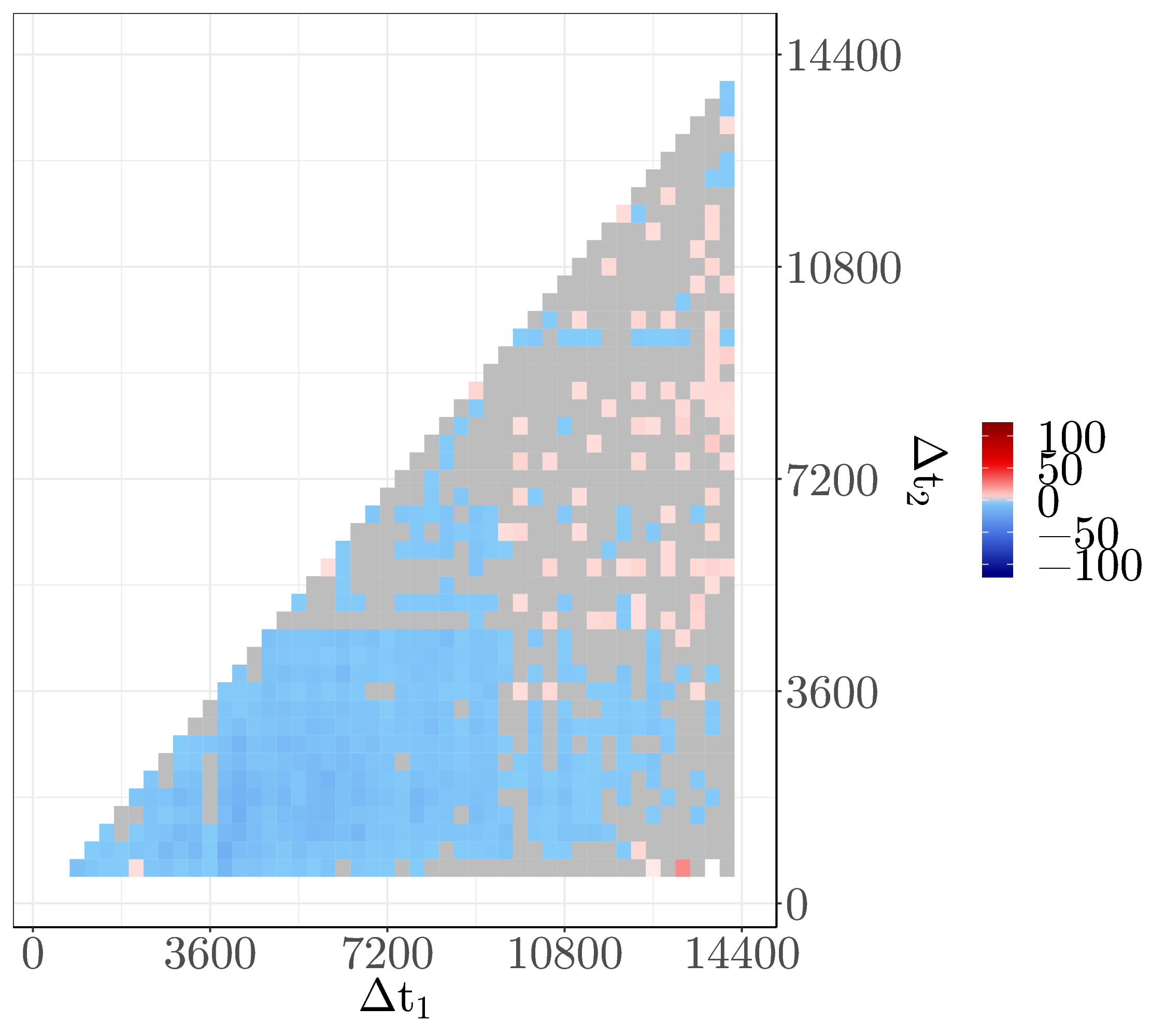}
\end{subfigure}

\caption{Left hand-side plot: average correlation between the leading (at timescale $\Delta t_1$) and lagging (at timescale $\Delta t_2$ activity rates, $\rho_N(\Delta t_1,\Delta t_2)$. Right hand-side plot: t-statistics of the difference $\rho_N(\Delta t_1,\Delta t_2)-\rho_N(\Delta t_2,\Delta t_1)$; negative values correspond to activity at small timescales being more correlated to future activity at larger timescale than reversely. Gray areas correspond to non-significant relationship according to a FDR of $0.05$. LB data set, $T_{\rm in}=120$.}
\label{fig:links_micro_cor_LB}
\end{figure}

\begin{figure}

\begin{subfigure}{.49\textwidth}

\centering
\caption{Average correlation}
\includegraphics[scale=0.35]
{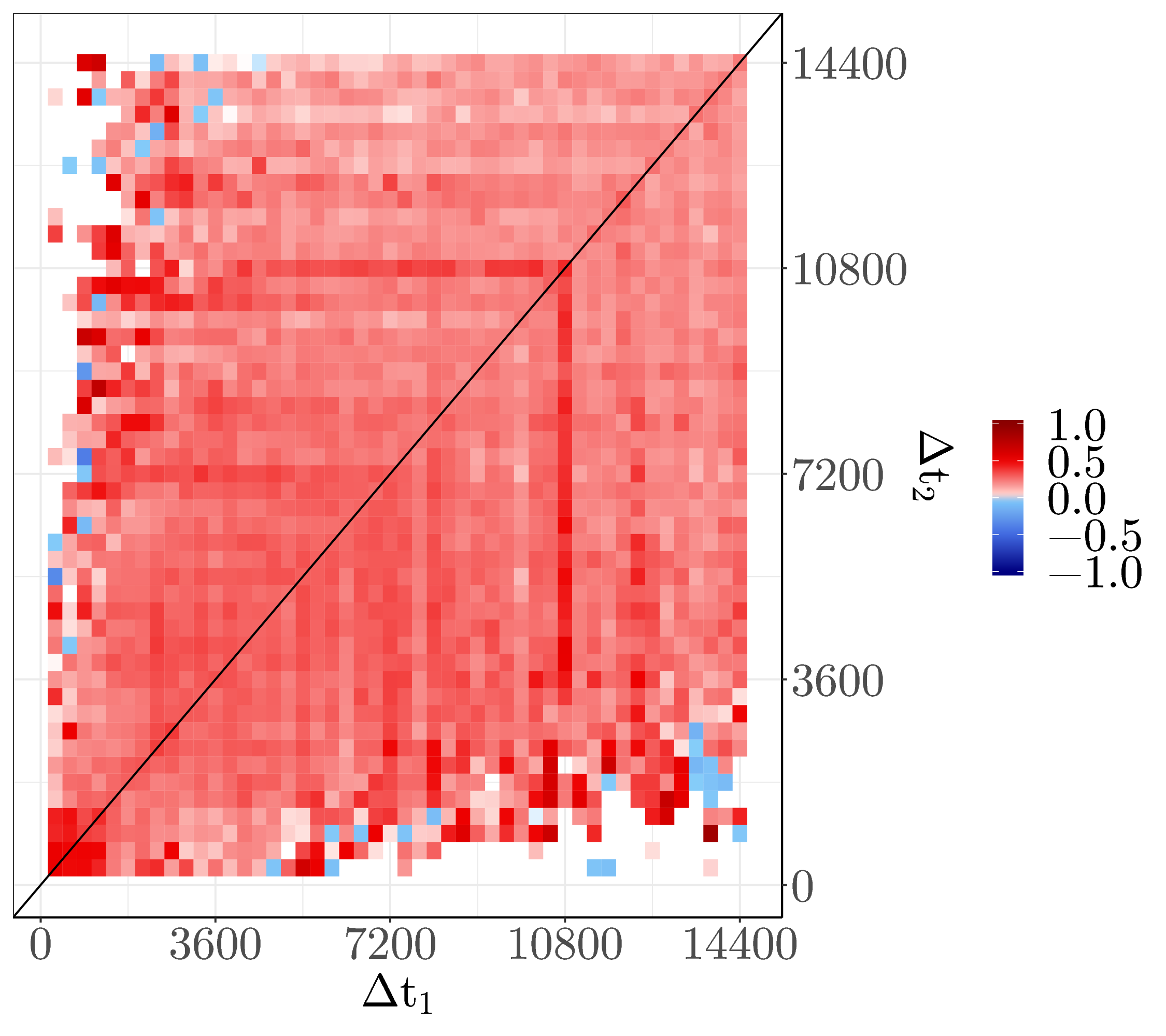}
\end{subfigure}
\begin{subfigure}{.49\textwidth}
\centering
\caption{FDR-corrected t-stat}
\includegraphics[scale=0.35]{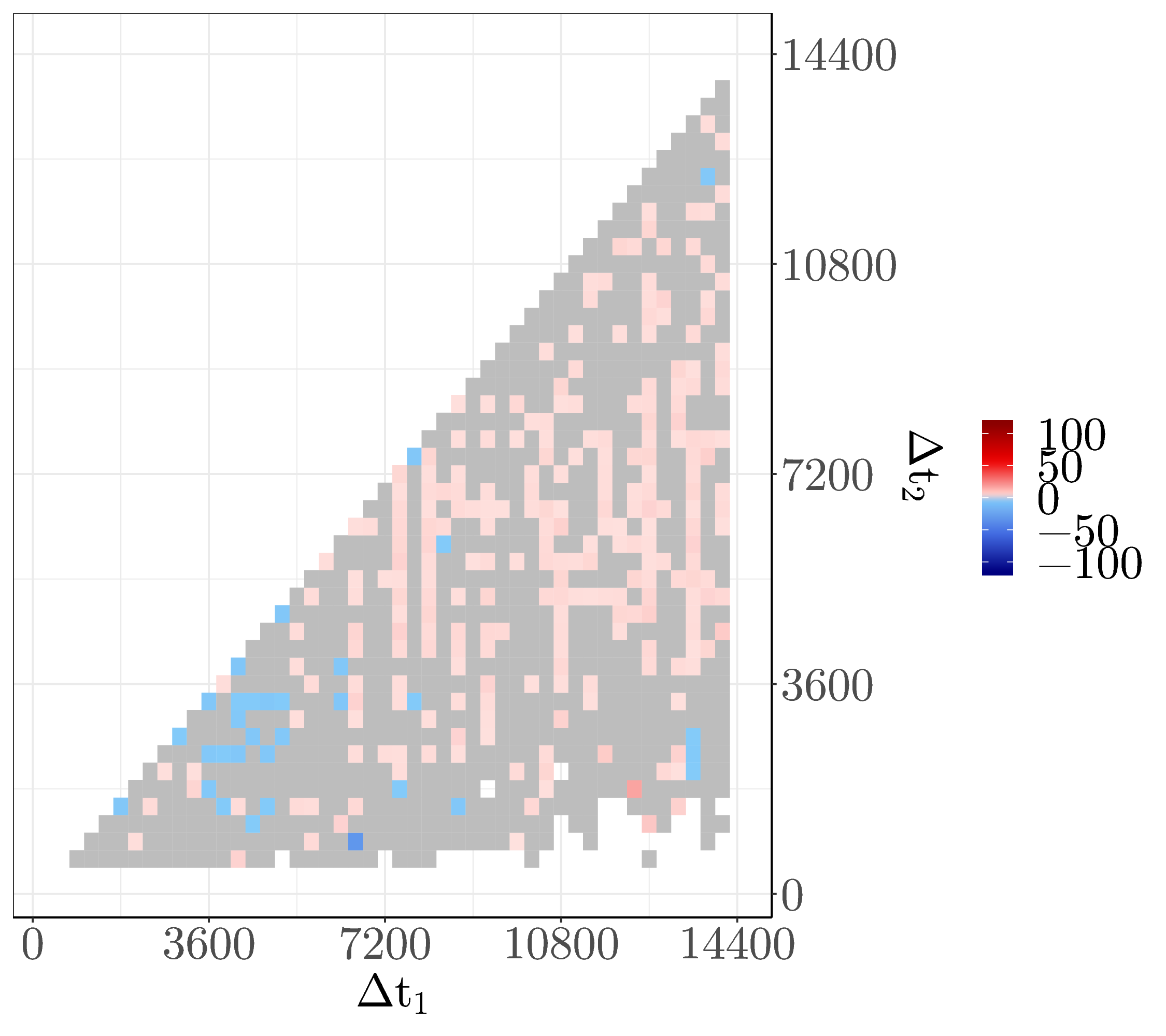}
\end{subfigure}

\caption{Left hand-side plot: average correlation between the leading (at timescale $\Delta t_1$) and lagging (at timescale $\Delta t_2$) activity rates, $\rho_N(\Delta t_1,\Delta t_2)$. Right hand-side plot: t-statistics of the difference $\rho_N(\Delta t_1,\Delta t_2)-\rho_N(\Delta t_2,\Delta t_1)$; negative values correspond to activity at small timescales being more correlated to future activity at larger timescale than reversely. Gray areas correspond to non-significant relationship according to a FDR of $0.05$. Gray areas correspond to non-significant relationship according to a FDR of $0.05$. SQ data set, $T_{\rm in}=60$ days.}
\label{fig:links_micro_cor_SQ}
\end{figure}

Since the lead-lag method introduced above specifically detects lead-lag activity, we expect to find clearer results if one focuses on statistically significant activity lead-lag. This is why we estimate the correlation between the activity rate of traders in leading groups and lagging groups, determined at two different timescales, as above.
Let us therefore denote the total number of trades of agents in group $1$ during time interval $[t-\Delta _1,t[$ by $N^{(1)}(t)$ and  the total number of trades of agents in group $2$ during time interval $[t,t+\Delta t_2[$ by $N^{(2)}(t)$. We then can compute the correlation between activity rates $N_1(t)/\Delta t_1$ and $N_2(t)/\Delta t_2$, denoted by $\rho_N(\Delta t_1,\Delta t_2)$. We perform this computation for each calibration window.

Figures~\ref{fig:links_micro_cor_LB} and \ref{fig:links_micro_cor_SQ} plot $\rho_N(\Delta t_1,\Delta t_2)$ for the LB and SQ data sets respectively (left-hand side plots), and correspond to the mugshots of \cite{zumbach2001heterogeneous} but for activity rates conditioned on the statistically significant lead-lag activity. The left plots report the lead-lags between activity rates, while the right plots display the t-statistics of the time series $\delta\rho_N(\Delta t_1,\Delta t_2)=\rho_N(\Delta t_1,\Delta t_2)-\rho_N(\Delta t_2,\Delta t_1)$ obtained in each calibration window; only the values validated by FDR at a 5\% threshold are in color, the not-validated ones being reported in grey.
 
Let us start with LB traders: the asymmetry is clear and is confirmed by the right-hand side plots.  Zumbach effect emerges (red zone) when $\Delta t_2 >5400s$ and $\Delta t_1>10800s$. At timescales shorter than $\Delta t_2 =5400s$ and $\Delta_1=10800s$, activity in the past is more correlated with future activity on longer timescales than the opposite (blue zone); this globally mirrors the dependence between the number of links and the correlation. 
 
Remarkably, the behaviour of SQ traders corresponds to the same TRA structure, albeit with much weaker anti-Zumbach effect for shorter timescales than for LB data: there is a small concentration of blue dots around $\Delta t_1=3600s$. This is consistent with the fact that typical time between trades of SQ traders is smaller than that of LB traders.

Our data set is not sufficiently long or dense to report results for $\Delta t_1$ or $\Delta t_2>14400s$. The comparison with the results of \cite{zumbach2001heterogeneous} is therefore difficult as they are about timescales larger than $14400s$. Even more, both \cite{zumbach2001heterogeneous} and \cite{zumbach2009time} do not use physical time, but business time, which makes comparisons even harder. However, in both cases, we do find TRA with methods inspired by these works.

\section{Conclusions}
\label{sec:conclusion}

Lead-lag SVNs between two timescales reveal a fine-grained picture of the relevant timescales and of the causal structure of activity in FX markets and more generally in complex systems. FX traders come in various sizes and varieties, which is reflected  in the different types of lead-lag SVNs we have reported here.
For example, the calibration window length at which our method detects most groups and links is much smaller for SQ retail clients, which also are more likely to have a  self-referential behavior than the institutional investors in the LB data set.

Trader activity partially propagates  on the lead-lag networks that our method infers. We chose to focus on signed activity, but the states can also be defined as active/inactive, which leads to another family of lead-lag networks. At any rate, two timescales are not enough to characterise market activity and feedback loops, which is why we had to consider 1176 pairs of timescales. The emerging picture of asymmetry between the mutual influence of timescales is compatible with the one reported in \cite{zumbach2001heterogeneous} and \cite{zumbach2009time} at pairs of timescales sufficiently large, even if the definitions of time and time coarsening are different in our case. At smaller timescales, both institutional and retail traders display an anti-Zumbach behaviour, although the anti-Zumbach effect is very weak for retail traders. 

We leave for future investigations the question of how Zumbach and anti-Zumbach effects are related to particular types of trading strategies used by traders, in a spirit similar to  \cite{farmer2002price}. Two main facts need to be explained: why both the behaviour of both LB and SQ traders lead to a Zumbach effect for sufficiently large timescales, and what kind of strategies cause or do not cause an anti-Zumbach effect at shorter timescales.

\bibliographystyle{abbrvnat}
\bibliography{my_references}

\begin{thebibliography}{27}
\providecommand{\natexlab}[1]{#1}
\providecommand{\url}[1]{\texttt{#1}}
\expandafter\ifx\csname urlstyle\endcsname\relax
  \providecommand{\doi}[1]{doi: #1}\else
  \providecommand{\doi}{doi: \begingroup \urlstyle{rm}\Url}\fi

\bibitem[Benjamini and Hochberg(1995)]{benjamini1995controlling}
Y.~Benjamini and Y.~Hochberg.
\newblock Controlling the false discovery rate: a practical and powerful
  approach to multiple testing.
\newblock \emph{Journal of the royal statistical society. Series B
  (Methodological)}, pages 289--300, 1995.

\bibitem[Borland and Bouchaud(2005)]{borland2005multi}
L.~Borland and J.-P. Bouchaud.
\newblock On a multi-timescale statistical feedback model for volatility
  fluctuations.
\newblock \emph{arXiv preprint physics/0507073}, 2005.

\bibitem[Boudoukh et~al.(1994)Boudoukh, Richardson, and
  Whitelaw]{boudoukh1994tale}
J.~Boudoukh, M.~P. Richardson, and R.~Whitelaw.
\newblock A tale of three schools: Insights on autocorrelations of
  short-horizon stock returns.
\newblock \emph{Review of financial studies}, 7\penalty0 (3):\penalty0
  539--573, 1994.

\bibitem[Challet et~al.(2018)Challet, Chicheportiche, Lallouache, and
  Kassibrakis]{challet2018trader}
D.~Challet, R.~Chicheportiche, M.~Lallouache, and S.~Kassibrakis.
\newblock Statistically validated lead-lag networks and inventory prediction in
  the foreign exchange market.
\newblock \emph{Advances in Complex Systems}, page 1850019, 2018.

\bibitem[Chicheportiche and Bouchaud(2014)]{chicheportiche2014fine}
R.~Chicheportiche and J.-P. Bouchaud.
\newblock The fine-structure of volatility feedback {I}: Multi-scale
  self-reflexivity.
\newblock \emph{Physica A: Statistical Mechanics and its Applications},
  410:\penalty0 174--195, 2014.

\bibitem[Clark(1973)]{clark1973subordination}
P.~K. Clark.
\newblock A subordinated stochastic process model with finite variance for
  speculative prices.
\newblock \emph{Econometrica}, 41\penalty0 (1):\penalty0 135--155, 1973.
\newblock ISSN 00129682, 14680262.
\newblock URL \url{http://www.jstor.org/stable/1913889}.

\bibitem[Dacorogna et~al.(1998)Dacorogna, M{\"u}ller, Pictet, and
  Olsen]{dacorogna1998modelling}
M.~Dacorogna, U.~M{\"u}ller, O.~Pictet, and R.~Olsen.
\newblock Modelling short-term volatility with \uppercase{GARCH} and
  \uppercase{HARCH} models.
\newblock 1998.

\bibitem[Farmer and Joshi(2002)]{farmer2002price}
J.~D. Farmer and S.~Joshi.
\newblock The price dynamics of common trading strategies.
\newblock \emph{Journal of Economic Behavior \& Organization}, 49\penalty0
  (2):\penalty0 149--171, 2002.

\bibitem[Ghashghaie et~al.(1996)Ghashghaie, Breymann, Peinke, Talkner, and
  Dodge]{ghashghaie1996turbulent}
S.~Ghashghaie, W.~Breymann, J.~Peinke, P.~Talkner, and Y.~Dodge.
\newblock Turbulent cascades in foreign exchange markets.
\newblock \emph{Nature}, 381\penalty0 (6585):\penalty0 767, 1996.

\bibitem[Gillemot et~al.(2006)Gillemot, Farmer, and Lillo]{gillemot2006there}
L.~Gillemot, J.~D. Farmer, and F.~Lillo.
\newblock There's more to volatility than volume.
\newblock \emph{Quantitative Finance}, 6\penalty0 (5):\penalty0 371--384, 2006.

\bibitem[Hommes(2006)]{hommes2006heterogeneous}
C.~H. Hommes.
\newblock Heterogeneous agent models in economics and finance.
\newblock \emph{Handbook of computational economics}, 2:\penalty0 1109--1186,
  2006.

\bibitem[Jegadeesh and Titman(1995)]{jegadeesh1995overreaction}
N.~Jegadeesh and S.~Titman.
\newblock Overreaction, delayed reaction, and contrarian profits.
\newblock \emph{The Review of Financial Studies}, 8\penalty0 (4):\penalty0
  973--993, 1995.

\bibitem[Kroujiline et~al.(2016)Kroujiline, Gusev, Ushanov, Sharov, and
  Govorkov]{kroujiline2016forecasting}
D.~Kroujiline, M.~Gusev, D.~Ushanov, S.~V. Sharov, and B.~Govorkov.
\newblock Forecasting stock market returns over multiple time horizons.
\newblock \emph{Quantitative Finance}, 16\penalty0 (11):\penalty0 1695--1712,
  2016.

\bibitem[Lancichinetti and Fortunato(2009)]{lancichinetti2009community}
A.~Lancichinetti and S.~Fortunato.
\newblock Community detection algorithms: a comparative analysis.
\newblock \emph{Physical review E}, 80\penalty0 (5):\penalty0 056117, 2009.

\bibitem[Li et~al.(2014)Li, Palchykov, Jiang, Kaski, Kert{\'e}sz, Miccich{\`e},
  Tumminello, Zhou, and Mantegna]{li2014statistically}
M.-X. Li, V.~Palchykov, Z.-Q. Jiang, K.~Kaski, J.~Kert{\'e}sz, S.~Miccich{\`e},
  M.~Tumminello, W.-X. Zhou, and R.~N. Mantegna.
\newblock Statistically validated mobile communication networks: the evolution
  of motifs in european and chinese data.
\newblock \emph{New Journal of Physics}, 16\penalty0 (8):\penalty0 083038,
  2014.

\bibitem[Lux et~al.(2001)]{lux2001turbulence}
T.~Lux et~al.
\newblock Turbulence in financial markets: the surprising explanatory power of
  simple cascade models.
\newblock \emph{Quantitative finance}, 1\penalty0 (6):\penalty0 632--640, 2001.

\bibitem[Lynch et~al.(2003)Lynch, Zumbach, et~al.]{lynch2003market}
P.~E. Lynch, G.~O. Zumbach, et~al.
\newblock Market heterogeneities and the causal structure of volatility.
\newblock \emph{Quantitative Finance}, 3\penalty0 (4):\penalty0 320--331, 2003.

\bibitem[Marsili and Piai(2002)]{marsili2002colored}
M.~Marsili and M.~Piai.
\newblock Colored minority games.
\newblock \emph{Physica A: Statistical Mechanics and its Applications},
  310\penalty0 (1-2):\penalty0 234--244, 2002.

\bibitem[Mosetti et~al.(2006)Mosetti, Challet, and Zhang]{mosetti2006minority}
G.~Mosetti, D.~Challet, and Y.-C. Zhang.
\newblock Minority games with heterogeneous timescales.
\newblock \emph{Physica A: Statistical Mechanics and its Applications},
  365\penalty0 (2):\penalty0 529--542, 2006.

\bibitem[M{\"u}ller et~al.(1993)M{\"u}ller, Dacorogna, Dav{\'e}, Pictet, Olsen,
  and Ward]{muller1993fractals}
U.~A. M{\"u}ller, M.~M. Dacorogna, R.~D. Dav{\'e}, O.~V. Pictet, R.~B. Olsen,
  and J.~R. Ward.
\newblock Fractals and intrinsic time: A challenge to econometricians.
\newblock \emph{Unpublished manuscript, Olsen \& Associates, Z{\"u}rich}, 1993.

\bibitem[M{\"u}ller et~al.(1997)M{\"u}ller, Dacorogna, Dav{\'e}, Olsen, Pictet,
  and Von~Weizs{\"a}cker]{muller1997volatilities}
U.~A. M{\"u}ller, M.~M. Dacorogna, R.~D. Dav{\'e}, R.~B. Olsen, O.~V. Pictet,
  and J.~E. Von~Weizs{\"a}cker.
\newblock Volatilities of different time resolutions—analyzing the dynamics
  of market components.
\newblock \emph{Journal of Empirical Finance}, 4\penalty0 (2-3):\penalty0
  213--239, 1997.

\bibitem[Musciotto et~al.(2018)Musciotto, Marotta, Piilo, and
  Mantegna]{musciotto2018long}
F.~Musciotto, L.~Marotta, J.~Piilo, and R.~N. Mantegna.
\newblock Long-term ecology of investors in a financial market.
\newblock \emph{Palgrave Communications}, 4\penalty0 (1):\penalty0 92, 2018.

\bibitem[Rosvall and Bergstrom(2008)]{rosvall2008maps}
M.~Rosvall and C.~T. Bergstrom.
\newblock Maps of random walks on complex networks reveal community structure.
\newblock \emph{Proceedings of the National Academy of Sciences}, 105\penalty0
  (4):\penalty0 1118--1123, 2008.

\bibitem[Tumminello et~al.(2011)Tumminello, Micciche, Lillo, Piilo, and
  Mantegna]{tumminello2011statistically}
M.~Tumminello, S.~Micciche, F.~Lillo, J.~Piilo, and R.~N. Mantegna.
\newblock Statistically validated networks in bipartite complex systems.
\newblock \emph{PloS one}, 6\penalty0 (3):\penalty0 e17994, 2011.

\bibitem[Tumminello et~al.(2012)Tumminello, Lillo, Piilo, and
  Mantegna]{tumminello2012identification}
M.~Tumminello, F.~Lillo, J.~Piilo, and R.~N. Mantegna.
\newblock Identification of clusters of investors from their real trading
  activity in a financial market.
\newblock \emph{New Journal of Physics}, 14\penalty0 (1):\penalty0 013041,
  2012.

\bibitem[Zumbach(2009)]{zumbach2009time}
G.~Zumbach.
\newblock Time reversal invariance in finance.
\newblock \emph{Quantitative Finance}, 9\penalty0 (5):\penalty0 505--515, 2009.

\bibitem[Zumbach and Lynch(2001)]{zumbach2001heterogeneous}
G.~Zumbach and P.~Lynch.
\newblock Heterogeneous volatility cascade in financial markets.
\newblock \emph{Physica A: Statistical Mechanics and its Applications},
  298\penalty0 (3-4):\penalty0 521--529, 2001.

\end{thebibliography}
\end{document}